\documentclass[
 reprint,
 amsmath,amssymb,
 aps,
prd,
]{revtex4-2}

\usepackage{xcolor}
\usepackage{graphicx}
\usepackage{dcolumn}
\usepackage{bm}
\usepackage[normalem]{ulem}

\begin{document}

\title{Squeezed Limit non-Gaussianity Estimation with Cosmic Shear}

\author{Shi-Hui Zang}
\affiliation{Department of Physics, University of Wisconsin-Madison, Madison, WI 53706, USA}

\author{Moritz M\"unchmeyer}
\affiliation{Department of Physics, University of Wisconsin-Madison, Madison, WI 53706, USA}

\date{\today}

\begin{abstract}
We present a new method to constrain local primordial non-Gaussianity using the large-scale modulation of the local lensing power spectrum. 
Our work extends our recently proposed $\pi$-field method for primordial non-Gaussianity estimation to spherical coordinates and applies it to galaxy lensing. Our approach is computationally efficient and only requires binned multipole power spectra $C_\ell(z_1,z_2)$ on large scales, as well as their covariance. 
Our method is simpler to implement than a full bispectrum estimator, but still contains the full squeezed-limit information. 
We validate our model using a suite of N-body simulations and demonstrate its accuracy in recovering the $f_{\mathrm{NL}}$ values. 
We then perform a Fisher forecast for an LSST-like weak lensing survey, finding $\sigma_{f_{\mathrm{NL}}} \simeq 44$. Our approach readily combines with other $f_{\mathrm{NL}}$-sensitive fields such as kSZ velocity reconstruction and clustering-based $\pi$-fields, for a future combined $f_{\mathrm{NL}}$ estimator using various large-scale galaxy and CMB observables. 
\end{abstract}

\maketitle


\section{Introduction}

Ongoing and upcoming sky surveys such as DESI \citep{DESI:2016fyo}, Euclid \citep{EuclidTheoryWorkingGroup:2012gxx}, SPHEREx \citep{SPHEREx:2014bgr}, and the Vera C. Rubin Observatory Legacy Survey of Space
and Time (LSST) \citep{LSSTScience:2009jmu} will provide unprecedented data about the distribution of matter in the universe. One of the key science goals of these surveys is to improve our understanding of cosmological inflation, which likely seeded the initial conditions of structure formation. 
While the single-field slow-roll inflation model predicts almost Gaussian initial conditions \citep{2003JHEP...05..013M, 2004JCAP...10..006C, 2011JCAP...11..038C, 2013PhRvD..88h3502P}, alternative theories often predict measurable deviations from Gaussianity \citep{2019BAAS...51c.107M, 2022arXiv220308128A}. 
Thus, precise estimation of primordial non-Gaussianity is essential for distinguishing between these models. 
The simplest and most widely studied form is local non-Gaussianity, parameterized by $f_{\mathrm{NL}}$, which is defined by a quadratic correction to the primordial gravitational potential field $\Phi_G(\bm{x})$:
\begin{equation}
    \Phi(\bm{x}) = \Phi_G (\bm{x}) + f_{\mathrm{NL}}\left(\Phi_G(\bm{x})^2 - \langle \Phi_G^2 \rangle \right) ,
    \label{eq:fnl_def}
\end{equation}
where $\Phi_G(\bm{x})$ is a Gaussian field, $\langle \cdot \rangle$ represents the ensemble mean. 
While single-field inflation models generally predict negligible local non-Gaussianity, multi-field models can give large signatures. A natural theoretical target (e.g. \cite{2019BAAS...51c.107M}) for local non-Gaussianity is $f_{\mathrm{NL}} \gtrsim 1$ for many multi-field inflation models. Upcoming large-scale structure surveys, in particular SPHEREx, will attempt to reach this goal, but will require excellent large-scale calibration of their galaxy data.

The current best constraint on $f_{\mathrm{NL}}$ comes from the Planck experiment's analysis of the cosmic microwave background (CMB), reaching $f_{\mathrm{NL}} = 0.9 \pm 5.1$ \citep{Planck:2019kim}. 
However, further improvements with CMB data are limited by cosmic variance and will not reach $\sigma_{f_{\mathrm{NL}}} = 1$. Consequently, attention has turned to future large-scale structure (LSS) surveys, which are expected to yield a tighter bound on $f_{\mathrm{NL}}$ (local) in the next years, using the method of scale-dependent bias 
\citep{2008PhRvD..77l3514D, 2008ApJ...677L..77M, 2008JCAP...08..031S, 2009MNRAS.396...85D}. Scale-dependent bias is a unique and promising method to constrain primordial non-Gaussianity in the squeezed limit, and will allow to beat CMB constraints substantially in the future (e.g. \citep{2021JCAP...12..049S}). 

While galaxy clustering provides a powerful probe of local primordial non-Gaussianity, cosmic shear fields from gravitational lensing offer complementary information. Unlike galaxy clustering, weak lensing (cosmic shear) fields lack a distinctive scale-dependent bias on large scales, since they probe the total matter distribution which is unbiased. However, $f_{\mathrm{NL}}$ can be extracted from higher-order statistics like the bispectrum and trispectrum.
Recently, \citet{2022PhRvD.106l3525G, 2024PhRvD.109d3515G} demonstrated that the squeezed bispectrum from cosmic shear fields from an LSST-like survey achieves a constraint of $\sigma_{f_{\mathrm{NL}}}\sim 16$. They used a non-perturbative model of the squeezed limit matter bispectrum, the LSS consistency relations from \cite{Peloso:2013zw,Esposito:2019jkb}. Higher-order statistics like the trispectrum contain additional information, though modeling these signals non-perturbatively is complex due to their numerous coefficients, large variance, and sensitivity to outliers. 
Other possible approaches to constrain $f_{\mathrm{NL}}$ with weak lensing include field-level inference \citep{2023MNRAS.520.5746A}, one-point analysis \citep{2020MNRAS.498..464F} and topological analysis \citep{2021JCAP...04..061B}.

In this work, we develop a method which is also sensitive to the squeezed limit bispectrum of weak lensing, but using the computationally simpler approach of \citep{2023PhRvD.107f1301G, 2023arXiv230503070G, 2szy-wypg}. \citet{2023arXiv230503070G} proposed a novel method to probe $f_{\mathrm{NL}}$ by leveraging the large-scale modulation of small-scale fields, there called ``$\pi$-fields". In \cite{2023arXiv230503070G} this method was applied to $\pi$-fields which are band-powers of the local position-dependent power spectrum, and in the present work we repeat this analysis for the case of weak lensing. In \cite{2023PhRvD.107f1301G}, it was shown that one can obtain even tighter constraints by learning an optimal $\pi$-field using a neural network, and we also plan to adapt this neural network approach to weak lensing in future work. The advantage of the ``$\pi$-fields" method over other approaches is that it is both computationally straight-forward and optimal for squeezed limit signatures. Even though our approach probes the primordial bispectrum, it is fully based on power spectra. For these power spectra, we can make a bias model which includes all nuisance parameters. Because the method is based on power spectra, the estimation of the covariance matrix is straight forward. Finally, as we will discuss below, the $\pi$-field formalism easily allows us to combine a large number of $f_{\mathrm{NL}}$-sensitive fields at power spectrum level, including CMB lensing or kSZ velocity reconstruction \cite{Munchmeyer:2018eey}. An added benefit of using multiple fields from different data sources, is that $f_{\mathrm{NL}}$ can be extracted from the cross-correlations only, reducing the problem of large-scale calibration systematics. 

This paper is organized as follows. Sec.~\ref{sec:Method} discusses the theoretical formalism for modeling the local power spectrum field in spherical coordinates. For both the matter field and the shear field, this leads to geometric factors that need to be included in the modeling. In Sec.~\ref{sec:application}, we validate our model with the Ulagam lensing simulations \citep{2024JCAP...03..062A}, first with matter fields and then with lensing convergence fields. Sec.~\ref{sec:fisher} presents a Fisher forecast for $f_{\mathrm{NL}}$ with an  LSST-like galaxy survey, finding consistent results with \cite{2024PhRvD.109d3515G}. We make comparison between the non-perturbative bispectrum from \cite{2024PhRvD.109d3515G}, projected to two dimensions with the exact geometric factor, and a simpler linear bias model in Sec. \ref{sec:linear_model}. We conclude with a discussion in Sec.~\ref{sec:discussion}.

\section{Formalism}
\label{sec:Method}

In this section we introduce the definition of the local power spectrum field $\pi(\bm{n})$ on the sphere and calculate its power spectrum using the non-perturbative model of the squeezed limit from \cite{2024PhRvD.109d3515G}.

\subsection{Field Definition and Conventions}

As explained in \cite{2023arXiv230503070G}, any field $\pi(\bm{x})$ which is sensitive to the locally measured primordial $\sigma_8(\bm{x})$ in a region of space will be sensitive to $f_{\mathrm{NL}}$ through its scale-dependent bias. This is because local non-Gaussianity $f_{\mathrm{NL}}$ is a large-scale modulation of small-scale power, so any field that is sensitive to local (primordial) power will have a large-scale modulation in an $f_{\mathrm{NL}}$ cosmology. In this work, we consider the simplest such field, the local (late-time) power spectrum. This follows \cite{2023arXiv230503070G} which developed the same approach for the matter field in a 3D box.

Since our goal is to apply the method to weak lensing, which is naturally performed in binned spherical coordinates, we first adapt the method to these coordinates. 

Specifically, for any 2D projected map $\delta(\bm{n})$ defined by its projection kernel $W^{\delta}(\chi)$ as
\begin{equation}
    \delta(\bm{n}) = \int_0^\infty W^{\delta}(\chi)\delta_m(\bm{n}, \chi) d\chi,
    \label{eq:projected_map}
\end{equation}
we define its associated local power spectrum field, $\pi_i^{\delta}(\bm{n})$, by first high-pass filtering $\delta(\bm{n})$ and then squaring it in pixel space:
\begin{equation}
    \pi_i^{\delta}(\bm{n}) = \left( \sum_{\ell = 0}^{\infty} \sum_{m = -\ell}^{\ell} \delta_{\ell m} W_i^{\mathrm{HP}}(\ell) Y_{\ell m}(\bm{n}) \right)^2.
    \label{eq:pi_def}
\end{equation}
Here, $\delta_{\ell m}$ are the spherical harmonic coefficients of $\delta(\bm{n})$, $Y_{\ell m}(\bm{n})$ are spherical harmonics, and $W_i^{\mathrm{HP}}(\ell)$ is a high-pass filter that selects a specific band of multipoles:
\begin{equation}
    W_i^{\mathrm{HP}}(\ell) = \begin{cases}
    1, & \text{if } \ \ell_{\mathrm{min}}^{i} < \ell < \ell_{\mathrm{max}}^{i}  \\
    0, & \text{otherwise}.
\end{cases}
\end{equation}
or a smoothed version thereof. The bounds $\ell_{\mathrm{min}}^{i}$ and $\ell_{\mathrm{max}}^{i}$ define the range of small-scale modes included in the $i$-th $\pi$ field.
In this work, depending on the window function, the projected density fluctuation field $\delta(\bm{n})$ could be either matter field at a certain redshift $\delta^m(\bm{n}, \chi_0)$ if the projection kernel is a delta function $W^\delta(\chi) = \delta(\chi - \chi_0)$ or top hat, or the convergence field $\kappa^i(\bm{n})$ if the projection kernel is lensing kernel $W^\delta(\chi) = W^{\kappa_g, (i)}(\chi)$.

\subsection{Relating the $\pi$ field power spectrum to the underlying matter power spectrum}
\label{sec:non-pertur}

We first calculate the cross-power spectrum of a $\pi$ field and a projected matter field $\delta$ in terms of the underlying matter power spectrum.  The multipole moments of a projected field $\delta(\bm{n})$ (Eq.~\eqref{eq:projected_map}) are given by:
\begin{equation}
    \delta_{\ell m} = 4\pi i^\ell \int_0^\infty  d\chi  W^\delta(\chi) \int \frac{d^3 \bm{k}}{(2\pi)^3}  j_\ell(k\chi)  Y_{\ell m}^\ast(\hat{\bm{k}})  \delta(\bm{k}, z).
\end{equation}
Here $j_\ell (k\chi)$ is the spherical Bessel function. 
For the local power spectrum field $\pi_{i}^{\delta}(\bm{n}) = [\delta_i^{\mathrm{HP}}(\bm{n})]^2$, where $\delta^{\mathrm{HP}}_{i, \ell m} = \delta_{\ell m}W_i^{\mathrm{HP}}(\ell)$, its multipole moments are related to those of the original field via the Gaunt integral \citep{2001PhRvD..64d3516C}:
\begin{equation}
\begin{aligned}
    \pi_{i, lm}^{\delta} & = \sum_{\ell_1\ell_2m_1m_2} \delta_{i, \ell_1m_1}^{\mathrm{HP}} \delta_{i, \ell_2m_2}^{ \mathrm{HP}}  \mathcal{G}_{m_1m_2m}^{\ell_1\ell_2\ell} \\
    & = \sum_{\ell_1\ell_2m_1m_2} W_i^{\mathrm{HP}}(\ell_1) W_i^{\mathrm{HP}}(\ell_2)  \delta_{\ell_1m_1}  \delta_{\ell_2m_2}  \mathcal{G}_{m_1m_2m}^{\ell_1\ell_2\ell}.
    \label{eq:pi_multipole}
\end{aligned}
\end{equation}
The cross-power spectrum between fields $\delta^a$ and $\pi_{i}^{\delta^b}$ is defined as $\langle \pi_{i, \ell m}^{\delta^b} \delta_{\ell m}^{a, \ast} \rangle = \delta_{\ell \ell'}\delta_{mm'} C_{\delta^a\pi_i^{\delta^b}}(\ell)$. Using Eq.~\eqref{eq:pi_multipole}, this can be expressed in terms of the angular bispectrum:
\begin{equation}
\begin{aligned}
    C_{\delta^a\pi_i^{\delta^b}}(\ell) =& \langle \pi_{i, \ell m}^{\delta^b} \delta_{\ell m}^{a, \ast} \rangle \\
    =& \sum_{\ell_1 \ell_2 m_1 m_2} \mathcal{G}_{m_1m_2m}^{\ell_1\ell_2\ell}  W_i^{\mathrm{HP}}(\ell_1)  W_i^{\mathrm{HP}}(\ell_2) \\ & \times \langle \delta_{\ell_1m_1}^{b}  \delta_{\ell_2m_2}^{b}  \delta_{\ell m}^{a,  \ast} \rangle.
\end{aligned}
\end{equation}

To accurately model the cross-power spectrum $C_{\delta^a\pi_{i}^{\delta^b}}(\ell)$ beyond the linear regime we employ a non-perturbative approach based on the squeezed-limit bispectrum from \cite{2024PhRvD.109d3515G}. We will comment below in Sec.~\ref{sec:linear_model} to the relation to a simpler biasing model. We thus relate the angular bispectrum to the underlying 3D bispectrum in the squeezed limit as follows:
\begin{widetext}
\begin{equation}
    \langle \delta_{\ell_1m_1}^{1}  \delta_{\ell_2m_2}^{2}  \delta_{\ell_3m_3}^{3} \rangle = \frac{8}{\pi^3}\mathcal{G}_{m_1m_2m_3}^{\ell_1\ell_2\ell_3} \int dr \ r^2 \left[ \prod_{i = 1}^{3}d\chi_i \ dk_i \  k_i^2W^{\delta^i}(\chi_i)j_{\ell_i}(k_i\chi_i)j_{\ell_i}(k_i r) \right]B(k_1, k_2, k_3, \chi_1, \chi_2, \chi_3).
\end{equation}
\begin{equation}
\begin{aligned}
    \langle \delta_{\ell_1m_1}^{1}  \delta_{\ell_2m_2}^{2}  \delta_{\ell_3m_3}^{3} \rangle \approx & \frac{2}{\pi}  \mathcal{G}_{m_1m_2m_3}^{\ell_1\ell_2\ell_3}  \int d\chi_1  d\chi_2  \frac{W^{\delta^1}(\chi_1)  W^{\delta^2}(\chi_2)  W^{\delta^3}(\chi_2)}{\chi_2^2} \\
    & \times \int dq  q^2  j_{\ell_1}(q\chi_1)  j_{\ell_1}(q\chi_2)  B\left( q, \frac{\ell_2 + 1/2}{\chi_2}, \frac{\ell_3 + 1/2}{\chi_2}; \chi_1, \chi_2, \chi_2 \right).
    \label{eq:3pt_2}
\end{aligned}
\end{equation}
\end{widetext}
The high-pass filter $W^{\mathrm{HP}}(\ell)$ enforce a squeezed limit condition with $\ell_2, \ell_3 \gg 100$. Meanwhile we only model the cross power spectrum on large scales thus $\ell_1 \equiv \ell_{\mathrm{soft}} < 100$.
Therefore it's safe to apply the Limber approximation for the spherical Bessel function $j_{\ell_2}(k_2\chi_2)$ and $j_{\ell_3}(k_3\chi_3)$. Under this approximation, we have $\ell_2 \simeq \ell_3$ and $\chi_2 \simeq \chi_3$. Thus we have Eq.~(\ref{eq:3pt_2}).
This leads to the final expression for the cross-power spectrum (see Appendix~\ref{sec:Gaunt} for details of the Gaunt integral summation):
\begin{equation}
\begin{aligned}
    C_{\delta^a\pi_i^{\delta^b}}(\ell) = &  \frac{1}{2\pi^2}  \sum_{\ell_1\ell_2}  (2\ell_1 + 1)(2\ell_2 + 1)  \begin{pmatrix} \ell_1 & \ell_2 & \ell \\ 0 & 0 & 0 \end{pmatrix}^2  \\ & \times W_i^{\mathrm{HP}}(\ell_1)  W_i^{\mathrm{HP}}(\ell_2) \\
    & \times \int d\chi_1  d\chi_2  \frac{W^{\delta^b}(\chi_2)^2  W^{\delta^a}(\chi_1)}{\chi_2^2} \\ & \times \int dq  q^2  j_\ell(q\chi_1)  j_{\ell}(q\chi_2) \\
    & \times B\left( q, \frac{\ell_1 + \frac{1}{2}}{\chi_1}, \frac{\ell_2 + \frac{1}{2}}{\chi_1}; \chi_1, \chi_2, \chi_2 \right).
    \label{eq:cl21_final}
\end{aligned}
\end{equation}

\subsection{Squeezed-limit matter bispectrum model}
To evaluate Eq.~\eqref{eq:cl21_final}, we require a model for the squeezed-limit bispectrum $B(q, k; \chi_q, \chi_k)$. 
Based on the consistency relation, which indicates that late-time gravitational evolution contributes terms scaling with positive powers of the soft mode $q$, while the $f_{\mathrm{NL}}$ contribution scales with negative powers of $q$, we adopt a polynomial ansatz:
\begin{equation}
\begin{aligned}
    B(q, k; \chi_q, \chi_k) = & a_{f_{\mathrm{NL}}}(k, \chi_q, \chi_k)  \frac{P(q, \chi_q)}{q^2 T(q)}  \\ & + \sum_{n=0}^{\infty} a_n(k, \chi_k)  \left( \frac{q}{k} \right)^n  P(q, \chi_q)  P(k, \chi_k).
    \label{eq:polynomial_ansatz}
\end{aligned}
\end{equation}
Here, $k$ represents the wavenumber of the hard mode. The first term arises from non-zero $f_{\mathrm{NL}}$, while the second term encapsulates non-linear gravitational evolution.
We adopt the transfer function $T(q)$ following \citet{Eisenstein:1997ik}, which is then normalized to unity on large scales.
In this work, we truncate the series at $n=0$, which provides sufficient accuracy for a wide range of soft modes \citet{2022PhRvD.106l3525G, 2024PhRvD.109d3515G}:
\begin{equation}
\begin{aligned}
    B(q, k; \chi_q, \chi_k) = & a_{f_{\mathrm{NL}}}(k, \chi_q, \chi_k)  P(k, \chi_k)  \frac{P(q, \chi_q)}{q^2 T(q)} \\ & + a_0(k, \chi_k)  P(k, \chi_k)  P(q, \chi_q).
    \label{eq:bispectrum_model_truncated}
\end{aligned}
\end{equation}
The coefficient $a_0(k, \chi_k)$ can be determined analytically from the consistency condition \citep{2014PhRvD..89l3522V, 2014PhRvD..90b3546N}:
\begin{equation}
    a_0(k, \chi_k) = 1 + \frac{13}{21}  \frac{\partial \log P(k, \chi_k)}{\partial \log D(\chi_k)} - \frac{1}{3}  \frac{\partial \log P(k, \chi_k)}{\partial \log k}.
    \label{eq:a_0_grav}
\end{equation}

The $f_{\mathrm{NL}}$ coefficient is modeled as \citep{2022PhRvD.106l3525G}:
\begin{equation}
    a_{f_{\mathrm{NL}}}(k, \chi_q, \chi_k) = \frac{6f_{\mathrm{NL}}\Omega_m H_0^2}{D(\chi_q)}  \frac{\partial \log P(k, \chi_k)}{\partial \log \sigma_8^2}.
    \label{eq:a_fnl}
\end{equation}

However, Eq.~\eqref{eq:a_0_grav} may break down on small scales and at low redshifts due to baryonic effects. Besides, non-zero $f_{\mathrm{NL}}$ not only gives rise to $a_{f_{\mathrm{NL}}}$ term, but also affects the amplitude of gravitational non-Gaussianity. 
To account for this effect, we introduce a normalization factor $A_0$ for gravitational non-Gaussianity, which is marginalized during parameter estimation. 
Thus, our final bispectrum model is:
\begin{equation}
\begin{aligned}
    B(q, k; \chi_q, \chi_k) = & a_{f_{\mathrm{NL}}}(k, \chi_q, \chi_k)  P(k, \chi_k)  \frac{P(q, \chi_q)}{q^2 T(q)} \\ & + A_0  a_0(k, \chi_k)  P(k, \chi_k)  P(q, \chi_q).
    \label{eq:bispectrum_model_final}
\end{aligned}
\end{equation}
In this parametrization the only free parameters are $f_{\mathrm{NL}}$ and $A_0$, so the parameter set directly corresponds to the $(f_{\mathrm{NL}},b_g)$ pair in the simplified large-scale bias model in Sec.~\ref{sec:linear_model}.

\subsection{Evaluating the Covariance Matrix}

We will base our $f_{\mathrm{NL}}$ estimates on the measured cross-power spectra $C_{\delta \pi}(\ell)$, using a power-spectrum-based likelihood (rather than mode-based). We thus need to evaluate the covariance matrix on these power spectra. The covariance matrix can be evaluated in the fiducial case of $f_{\mathrm{NL}}=0$ and assuming Gaussianity of the fields (since we are on large scales only). 
The expression for the cross-power spectrum between two projected fields $\delta^a(\bm{n})$ and $\delta^b(\bm{n})$, which is necessary for the covariance estimation is given by
\begin{equation}
\begin{aligned}
    C_{\delta^a\delta^b}(\ell) = &  \frac{2}{\pi}\int W^{\delta^a}(\chi_1)d\chi_1 \int W^{\delta^b}(\chi_2)d\chi_2
 \\ & \times \int dq q^2j_{\ell}(q\chi_1)j_\ell(q\chi_2)P(q, \chi_1, \chi_2) \\ & + N_{\delta^a\delta^b}(\ell).
    \label{eq:Power_Accu}
\end{aligned}
\end{equation}
Here $P(q, \chi_1, \chi_2)$ is the non-equal-time matter power spectrum. 
The noise term $N_{\delta^a\delta^b}(\ell)$ is non-zero for the convergence field, where it corresponds to the shape noise arising from the intrinsic ellipticities of galaxies:
\begin{equation}
    N_{\kappa^i \kappa^j}(\ell) =  \delta^K_{ij} \frac{\sigma_\epsilon^2}{\bar{n}_g^{(i)}}.
\end{equation}
We adopt an effective number density of galaxies $\bar{n}_g = 31 \ \mathrm{arcmin}^{-2}$ and an intrinsic rms ellipticity $\sigma_\epsilon = 0.26$ following the LSST survey specifications \citep[][]{Chang:2013xja}.
In the case of multiple tomographic bins, the effective number density in each bin is $\bar{n}_g^{(i)} = \bar{n}_g / N_{\mathrm{tomo}}$.

The auto-power spectrum of the $\pi$ field,  assuming Gaussianity of the underlying fields, is given by:
\begin{equation}
\begin{aligned}
    C_{\pi_i^{\delta^a}\pi_j^{\delta^b}}(\ell) = &  \langle \pi_{i, \ell m}^{\delta^a}  \pi_{j, \ell'm'}^{\delta^b, \ast} \rangle \\
    = &  \sum_{\ell_1  \ell_2}   \frac{(2\ell_1 + 1)(2\ell_2 + 1)}{2\pi}  \begin{pmatrix} \ell_1 & \ell_2 & \ell \\ 0 & 0 & 0 \end{pmatrix}^2 \\ & \times W_i^{\mathrm{HP}}(\ell_1)  W_i^{\mathrm{HP}}(\ell_2)  W_{j}^{\mathrm{HP}}(\ell_1)  W_{j}^{\mathrm{HP}}(\ell_2) \\
    & \times C_{\delta^a\delta^b}(\ell_1)  C_{\delta^a\delta^b}(\ell_2).
    \label{eq:pi_auto}
\end{aligned}
\end{equation}

Finally, with the Gaussian approximation and Wick's theorem, the covariance between two cross power spectra can be derived as follows:
\begin{widetext}
\begin{equation}
\begin{aligned}
    \mathrm{Cov}\left[ C_{\delta^a\pi_i^{\delta^b}}(\ell), C_{\delta^{c}\pi_{j}^{\delta^d}}(\ell') \right]  = &  \langle C_{\delta^{a}\pi_i^{\delta^b}}(\ell)C_{\delta^{c}\pi_j^{\delta^d}}(\ell')\rangle - \langle C_{\delta^{a}\pi_i^{\delta^b}}(\ell) \rangle \langle C_{\delta^{c}\pi_j^{\delta^d}}(\ell')\rangle \\
     = &  \langle \pi_{i, \ell m}^{\delta^b}\delta_{\ell m}^{a, \ast} \pi_{j, \ell' m'}^{\delta^d} \delta_{\ell' m'}^{c, \ast} \rangle - \langle \pi_{i, \ell m}^{\delta^b}\delta_{\ell m}^{a, \ast} \rangle \langle \pi_{j, \ell' m'}^{\delta^d} \delta_{\ell' m'}^{c, \ast}\rangle \\
     = &  \langle\pi_{i, \ell m}^{\delta^b} \delta_{\ell m}^{a, \ast}\rangle\langle\pi_{j, \ell' m'}^{\delta^d}\delta_{\ell' m'}^{c, \ast}  \rangle + \langle \pi_{i, \ell m }^{\delta^b}\pi_{j, \ell' m'}^{\delta^d} \rangle \langle \delta_{\ell m}^{a, \ast} \delta_{\ell' m'}^{c, \ast} \rangle \\ &  + \langle \pi_{i, \ell m}^{\delta^b} \delta_{\ell' m'}^{c, \ast}\rangle\langle \delta_{\ell m}^{a, \ast} \pi_{j, \ell' m'}^{\delta^d}\rangle - \langle \pi_{i, \ell m}^{\delta^b}\delta_{\ell m}^{a, \ast} \rangle \langle \pi_{j, \ell' m'}^{\delta^d} \delta_{\ell' m'}^{c, \ast}\rangle \\
     = &  \frac{\delta_{\ell\ell'}}{(2\ell + 1)f_{\mathrm{sky}}}\left[ C_{\delta^a \pi_j^{\delta^d}}(\ell)C_{\delta^{c} \pi_i^{\delta^b}}(\ell) + C_{\delta^a\delta^{c}}(\ell) C_{\pi_i
     ^{\delta^b}\pi_j^{\delta^d}}(\ell) \right], \label{eq:Cov}
\end{aligned} 
\end{equation}
\end{widetext}
with $f_{\mathrm{sky}}$ referring to sky coverage of the survey. 

\section{Application to the Ulagam Simulations}
\label{sec:application}
In this section, we will verify our model using high-resolution N-body simulations. We begin by applying our formalism to the matter field and then extend the results to the lensing convergence fields.

\subsection{Simulations}

Several recent suites of high-resolution N-body simulations, including Quijote \citep{2020ApJS..250....2V} and AbacusSummit \citep{2021MNRAS.508.4017M}, allow to test high-order summary statistics in detail. Unfortunately, our analysis has demanding simulation requirements. First, to test unbiasedness of our $f_{\mathrm{NL}}$ estimate, we need simulations with $f_{\mathrm{NL}} \neq 0$. Second, we need light-cone simulations of cosmic shear. The only available simulation set that meets these two requirements is the Ulagam simulation \citep{2024JCAP...03..062A}. 

The Ulagam simulation set consists of over 4,000 simulations with varying cosmological parameters, run by Pkdgrav3 solver \citep{2017ComAC...4....2P}.
Each simulation outputs lightcone dark matter fields at 100 different redshifts from initial condition at $z = 127$ to final condition at $z = 0$. These density fields are stored in the form of number counts of dark matter particles in ${\tt HEALPix}$ \citep{2005ApJ...622..759G} pixels with ${\tt nside} = 1024$. We compute the matter density contrast field as:
\begin{equation}
    \delta_m(\bm{n}) = N_{\mathrm{p}}(\bm{n}) / \langle N_p \rangle - 1.
\end{equation}
Here $N_{\mathrm{p}}(\bm{n})$ is the number of dark matter particles in a pixel, $\langle N_p \rangle$ is the average number of particles in each pixel.

For our study of $f_{\mathrm{NL}}$ signal in $\pi$ fields, we need both simulations with Gaussian and non-Gaussian initial conditions.
These correspond to three subsets of Ulagam simulation {\tt fiducial}, {\tt LC\_p} and {\tt LC\_m} as shown in Table~\ref{table:Ulagam}. Every subset is made up of 100 simulations. The ${\tt fiducial}$ set is simulated with cosmological parameter set ${\Omega_m = } 0.3175, \Omega_b = 0.049, h = 0.6711, n_s = 0.9624, w = -1, \sigma = 0.834$ and $f_{\mathrm{NL}} = 0$. 
The ${\tt LC\_p}$ subset is of same configuration except for non-zero local non-Gaussianity $f_{\mathrm{NL}} = 100$, while ${\tt LC\_m}$ is simulated with $f_{\mathrm{NL}} = -100$. The public library ${\tt healpy}$ \citep{Zonca2019} is used to read the density files and calculate the angular power spectra of fields.

\begin{table*}[]
\centering
\caption{The configuration of the subset of the Ulagam simulations used in this work.}
\label{table:Ulagam}
\begin{tabular}{ccccccccc}
\hline
Simulation      & $\Omega_m$                          & $\Omega_b$                         & $h$                                 & $n_s$                               & $\sigma_8$                         & $M_{\nu}$(eV)                            &  $f_{\mathrm{NL}}$ & Realizations \\ \hline
{\tt fiducial}        & 0.3175                              & 0.049                              & 0.6711                              & 0.9624                              & 0.834                              & 0.0                           & 0        & 100           \\ {\tt LC\_p}        & 0.3175                              & 0.049                              & 0.6711                              & 0.9624                              & 0.834                              & 0.0                           & 100        & 100           \\ {\tt LC\_m}        & 0.3175                              & 0.049                              & 0.6711                              & 0.9624                              & 0.834                              & 0.0                           & -100        & 100           \\ \hline
\end{tabular}
\end{table*}

Unfortunately, the Ulagam simulations have some limitations, since they were not designed for scale-dependent bias. To make fullsky maps, the simulation box (with side length $L = 1\,h^{-1}\,\mathrm{Gpc}$) is tiled to construct the lightcones, leading to biased large-scale information. For more details about this tiling effect please refer to Appendix~\ref{sec:Tiling}. In this paper, we will only use this simulation suite to verify our model at low redshifts where the tiling effect is small. While the simulation tiling limits our ability to use these simulations for a precision test, we still found them valuable to debug and test our pipeline. In particular, for cosmic shear one cannot entirely eliminate non-linear physics with a scale cut, since the lensing kernel has contributions from small redshifts where even large angles correspond to non-linear comoving $k$. This effect is included in the simulations, but would be hard to model analytically.

\subsection{Generating the $\pi$ fields and evaluating the likelihood}

For each projected map $\delta$ (either matter or weak lensing convergence fields), we use Eq.~\eqref{eq:pi_def} to generate one $\pi$ field $\{ \pi_1 ^ {\delta}\}$ with high-pass filter parameters $(\ell_{\mathrm{min}}^{1}, \ell_{\mathrm{max}}^{1}) = ( 500, 510)$. This choice is made to validate our power spectrum model on the Ulagam simulations at low redshift (to avoid the tiling regime of the simulations), and not to exhaustively extract $f_{\mathrm{NL}}$ information. We will use higher redshifts, wider high-pass filters, and multiple $\ell$ windows below in our Fisher forecast, where the goal is to set the tightest possible constraints.

We have two free parameters in our model: the polynomial coefficient $A_0$ and the primordial non-Gaussianity parameter $f_{\mathrm{NL}}$. To  estimate $f_{\mathrm{NL}}$ and to marginalize over ${A}_0$, we will run  Monte Carlo Markov Chain sampling on the estimated power spectra from the simulations. We construct a Gaussian likelihood function for the cross power spectrum.
\begin{equation}
\begin{aligned}
    \log\tilde{\mathcal{L}} = & -\frac{1}{2}\left[ C_{\delta^a\pi_i^{\delta^b}}(\ell) - C_{\delta_a\pi_i^{\delta^b}}^{\mathrm{model}} (\ell)\right]^T \\ & \times \mathrm{Cov}^{-1}\left[C_{\delta_a\pi_i^{\delta^b}}(\ell), C_{\delta_c\pi_j^{\delta^d}}(\ell')\right] \\ & \times \left[ C_{\delta^c\pi_j^{\delta^d}}(\ell') - C_{\delta_c\pi_j^{\delta^d}}^{\mathrm{model}} (\ell')\right].
\end{aligned}
\end{equation}
The above likelihood $\tilde{\mathcal{L}}$ is sampled with MCMC algorithm ${\tt emcee}$ \citep{2013PASP..125..306F} to obtain the estimation of $A_0$ and $f_{\mathrm{NL}}$. We have opted here for a power spectrum level likelihood, rather than a mode-based likelihood, to isolate the contribution from the squeezed-limit bispectrum. As shown in \citet{2023arXiv230503070G}, one can also include the squeezed limit trispectrum by including $C_{\pi_a\pi_i}$ but we leave this to future work to avoid modeling the non-perturbative trispectrum. 

\subsection{Application to the matter field}

In this section, we validate our model with matter fields at redshift $z=0.23$. This corresponds to the special case when the window function is a delta function $W(\chi) = \delta(\chi - \chi_0)$, where $\chi_0$ is the comoving distance at a specific light-cone shell. As explained above, we use this low redshift to avoid biases from the tiling of simulation cubes in Ulagam. We further enforce $\ell\geq5$, which is roughly the multipole scale at which tiling becomes relevant for a $1$ Gpc base simulation box.

\begin{figure*}[ht!]
    \centering    \includegraphics[width=\textwidth]{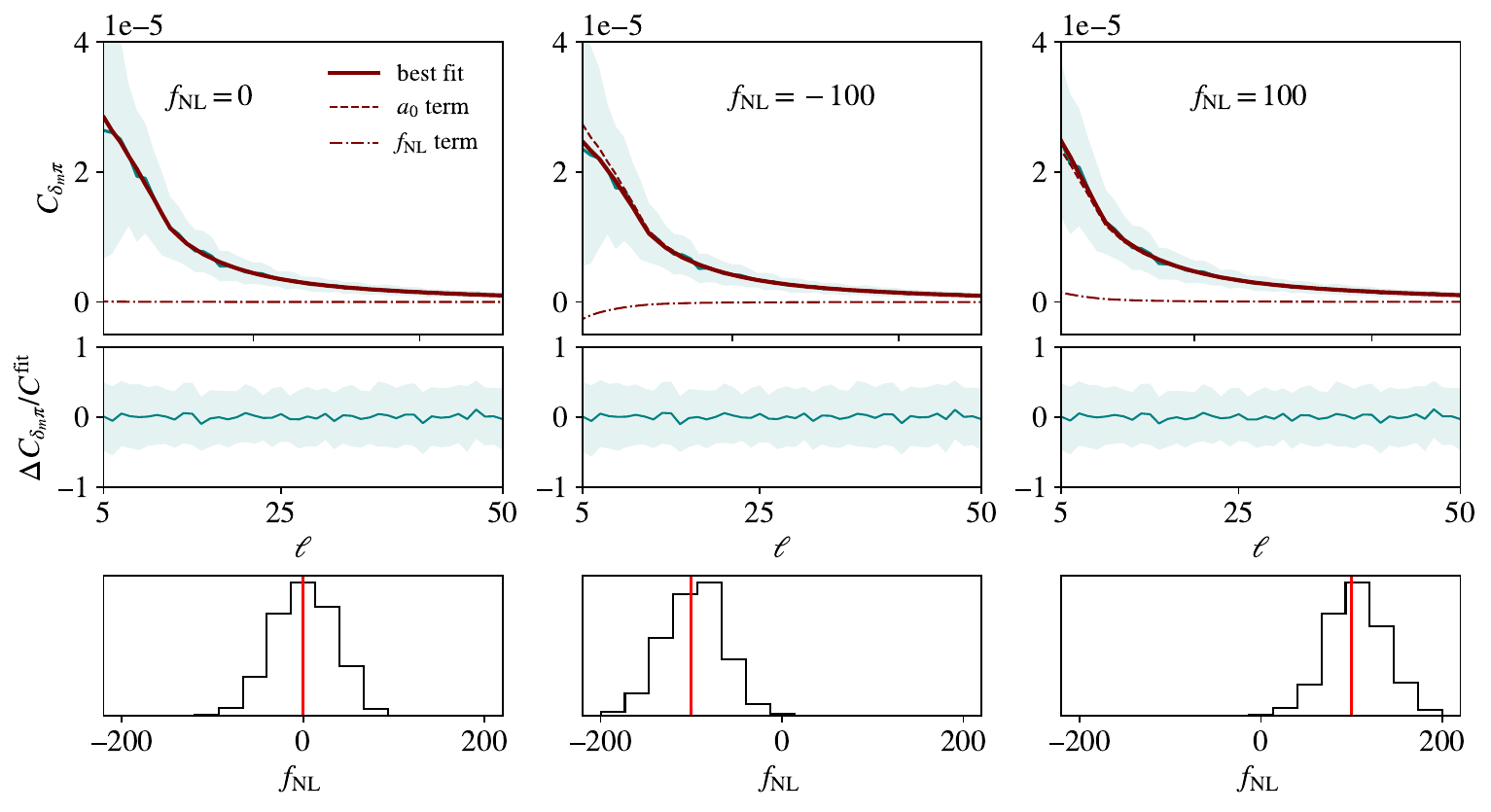}
    \caption{\textbf{Matter field model.} \textbf{Upper panel}: The blue curves show the cross-power spectrum $C_{\delta_m\pi}(\ell)$ for matter fields and $\pi$ fields from Ulagam simulations. The brown lines represent our best-fit models, with the dashed lines showing the contributions from the different terms in our model. The left, middle, and right panels correspond to cosmologies with input $f_{\mathrm{NL}}=0$, $f_{\mathrm{NL}}=100$, and $f_{\mathrm{NL}}=-100$, respectively. The redshift for the matter field is $z = 0.23$, and the high-pass filter $W^{\mathrm{HP}}$ filters out spherical harmonics within $(500, 510)$. \textbf{Middle panel}: The blue curves show the relative error of the simulated cross power spectrum compared with the modeled results. \textbf{Lower Panel}: The MCMC result of $f_{\mathrm{NL}}$ value after marginalizing $A_0$.}
    \label{fig:Matter_fit_plot2}
\end{figure*}

In Fig.~\ref{fig:Matter_fit_plot2} we plot the simulated cross-power spectrum and our modeling result for the matter field, with error bars estimated from simulations. 
The left, middle and right panels shows the modeling results of cosmologies with $f_{\mathrm{NL}} = 0, \  100, \ -100$. All results are averaged over 100 simulations. 
As is shown, the simulated result agrees with the model well. 
The gravitational non-Gaussianity term contributes the most to the cross spectrum, while $f_{\mathrm{NL}}$ term contributes only a small fraction of the total cross-power spectrum. This is consistent with \citet{2022PhRvD.106l3525G} where the 0-th order terms dominates over other higher order terms in the squeezed-limit bispectrum.

The lower panel of Fig.~\ref{fig:Matter_fit_plot2} shows the $f_{\mathrm{NL}}$ constraints from an MCMC analysis after marginalizing over $A_0$.
As expected, non-zero $f_{\mathrm{NL}}$ gives rise to the $f_{\mathrm{NL}}$ term (dash-dotted line), while in the fiducial case, $f_{\mathrm{NL}}$ is consistent with zero. 
Moreover, changing the $f_{\mathrm{NL}}$ value will also slightly affect the coefficients $A_0$. 
Numerically, for the fiducial cosmology we get $f_{\mathrm{NL}} = 3^{+31}_{-33}$, and $f_{\mathrm{NL}} = 105^{+31}_{-31}$, $f_{\mathrm{NL}} = -101^{+33}_{-32}$ for $f_{\mathrm{NL}} = 100, \ -100$ respectively. 

\begin{figure}[ht!]
    \centering    \includegraphics[width=\columnwidth]{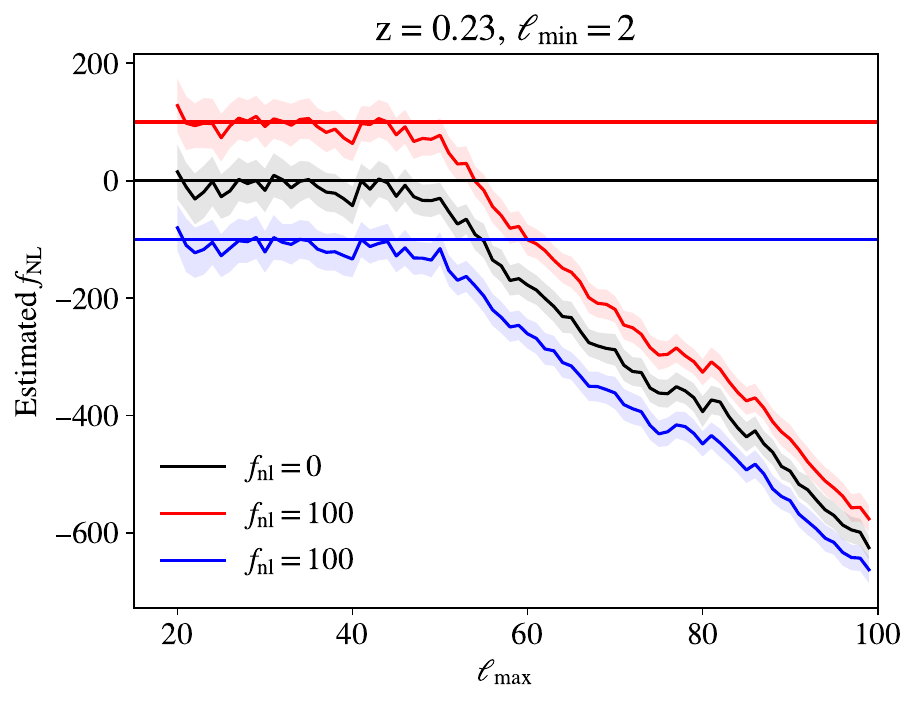}
    \caption{\textbf{Range of validity of the model for matter.} $f_{\mathrm{NL}}$ constraints as a function of $\ell_{\mathrm{max}}$. The black curve shows the result from $f_{\mathrm{NL}} = 0$ simulation subset, the red curve for $f_{\mathrm{NL}} = 100$ subset and blue curve for $f_{\mathrm{NL}} = -100$ subset. Shaded area shows the 1$\sigma$ confidence interval. The matter field shell is at redshift $z = 0.23$ and $\ell\geq 2$. }
    \label{fig:lmax}
\end{figure}

Fig.~\ref{fig:lmax} shows the estimated $f_{\mathrm{NL}}$ as a function of $\ell_{\mathrm{max}}$. The shaded area corresponds to the constraints. Generally by adding modes we're extracting more information from cross power spectrum. After $\mathrm{\ell_{\mathrm{max}}} > 45$, the estimated $f_{\mathrm{NL}}$ starts to deviate from true value significantly. At these scales, higher order terms in bispectrum become relevant, and the squeezed limit approximation starts to break down. A more general bispectrum model would be needed to push our method into smaller scales to retain more $f_{\mathrm{NL}}$ information. 

\subsection{Application to cosmic shear field}
In the previous section, we evaluate the model based on the $\pi$ field derived from matter shells. However, in a practical weak lensing survey, the matter field is not a directly observable. 
Instead, we observe the convergence field $\kappa_g$ (derived from cosmic shear) from galaxy shapes.
This convergence field $\kappa_g$ is a weighted projection of the matter field with following window function,
\begin{equation}
W^{\kappa_g^{(i)}}(\chi) = \frac{3 H_0^2 \Omega_m}{2 a(\chi)} \int_{\chi}^{\infty} d\chi' p_s^{(i)}(\chi') \left(\frac{\chi' - \chi}{\chi'}\right),
\label{eq:galaxy_lensing_window}
\end{equation}
where $p_s^{(i)}(\chi)$ is the redshift distribution function of galaxies at comoving distance $\chi$ in the $i$-th tomographic bin and $a(\chi)$ is the scale factor at comoving distance $\chi$. 
In this work we use $p_s^{(i)}(\chi)$ from LSST configuration.
For more details about $p_s^{(i)}(\chi)$ please refer to Sec.~\ref{sec:Suvey_Specifications} below. 

\begin{figure*}[ht!]
    \centering    \includegraphics[width=\textwidth]{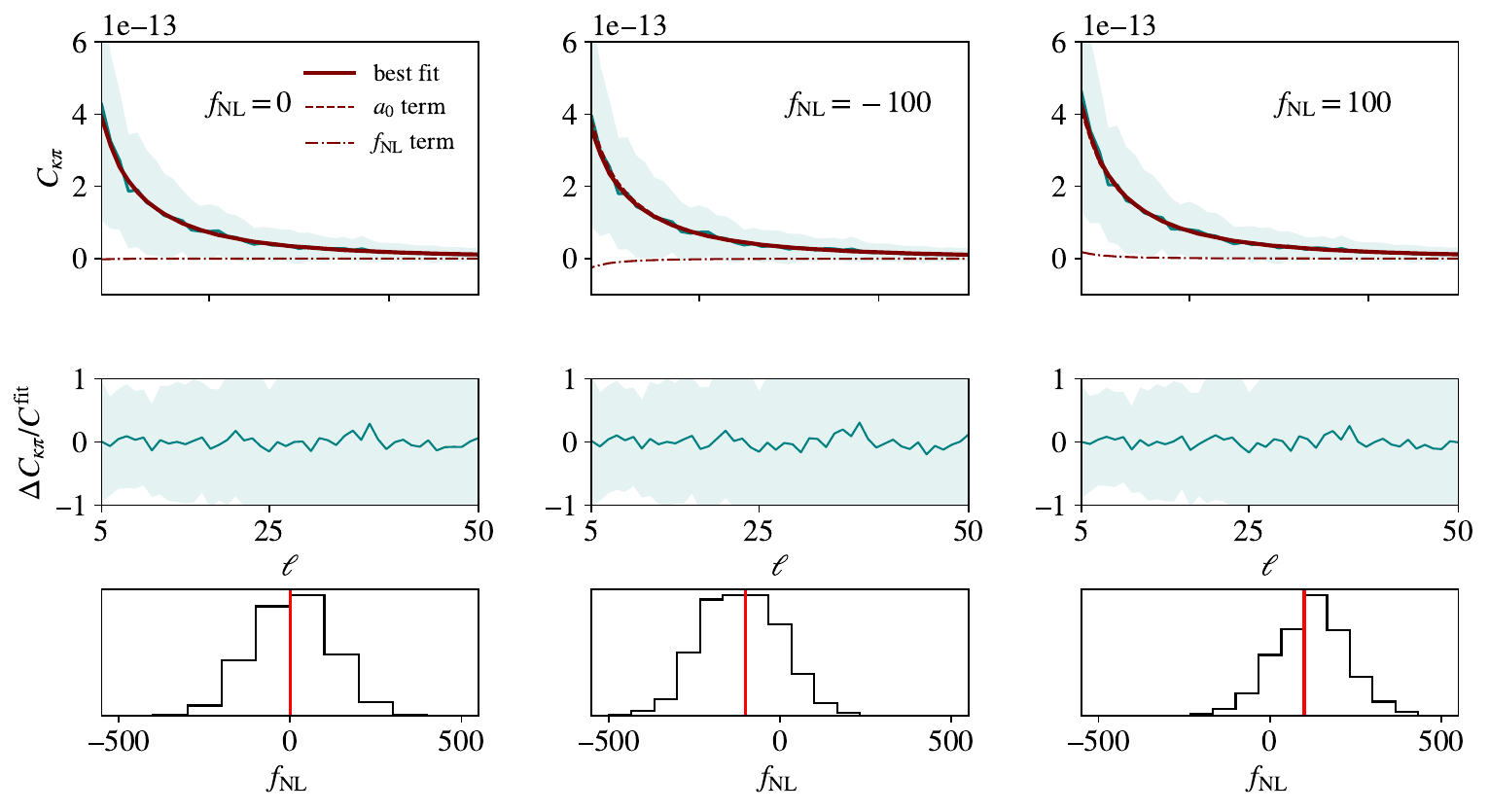}
    \caption{\textbf{Lensing field model in first of $N_{\mathrm{tomo}} = 3$ case.} \textbf{Upper panel}: The blue curves show the cross-power spectrum $C_{\kappa_g\pi}(\ell)$ for lensing convergence fields and $\pi$ fields from Ulagam simulations (using the first bin of three tomographic bins, with $\ell>5$). The brown lines represent our best-fit models, with the dashed lines showing the contributions from the different terms in our model. The left, middle, and right panels correspond to cosmologies with input $f_{\mathrm{NL}}=0$, $f_{\mathrm{NL}}=100$, and $f_{\mathrm{NL}}=-100$, respectively. The error bars are estimated from the scatter across 100 simulations. Here we assume the first of three tomographic bins, and the high-pass filter $W^{\mathrm{HP}}$ filters out spherical harmonics within $(500, 510)$. \textbf{Middle panel}: The blue curves show the relative error of the simulated cross power spectrum compared with the modeled results. \textbf{Lower Panel}: The MCMC result of $f_{\mathrm{NL}}$ value after marginalizing $a_0$.}
    \label{fig:Kappa_fit_plot2}
\end{figure*}

In the upper panel of Fig.~\ref{fig:Kappa_fit_plot2}, we plot the simulated cross power spectrum of the lensing convergence field $\kappa_g$ and its associated $\pi$ field, together with the modeled result. 
Both the simulated result and error bars are obtained from 100 simulations.  
The $\pi$ field is generated with a high-pass filter which filters out modes within $\ell \in [500, 510]$.
On large scale at $ \ell \leq 50$, the model agrees well with the simulation for different $f_{\mathrm{NL}}$ values.

In the lower panel of Fig.~\ref{fig:Kappa_fit_plot2}, we show the MCMC result of $f_{\mathrm{NL}}$ estimation with spherical harmonics $5 \leq \ell \leq 20$. Our model gives unbiased estimation of $f_{\mathrm{NL}}$ in all three cases. 
The numerical results are $f_{\mathrm{NL}} = 10^{113}_{-109}$, $f_{\mathrm{NL}} =-116^{+117}_{-109}$ and  $f_{\mathrm{NL}} = 124^{+116}_{-104}$ for $f_{\mathrm{NL}} = 0, -100, 100$ respectively.
Adding more modes and increasing $\ell_{\mathrm{max}}$ will further tighten this constraint. However, high-order terms becomes non-negligible and will bias the $f_{\mathrm{NL}}$ estimation.

Note that here we only show the modeling result of the lowest redshift bin of $N_{\mathrm{tomo}} = 3$ case, here $N_{\mathrm{tomo}}$ refers to the total number of tomographic bins (for more details about tomographic configuration please refer to Sec.~\ref{sec:Suvey_Specifications}).
We choose this bin for illustration because it is least affected by the tiling effect and provides the relatively cleaner validation case.

\section{Fisher Forecast for LSST experiment}
\label{sec:fisher}

In this section, we will evaluate the Fisher information of $f_{\mathrm{NL}}$ contained in the cross power spectrum $C_{\kappa_g \pi}$  under the LSST configuration. 
This forecast is based on the theoretical modeling of the power spectra and their covariance, and is not affected by the simulation tiling discussed above.

\subsection{Survey Specifications}
\label{sec:Suvey_Specifications}
The lensing convergence fields are constructed with the same redshift distribution model as \citet{2024PhRvD.109d3515G} given by LSST Dark Energy Science Collaboration Science Requirements Document \citep{LSSTDarkEnergyScience:2018jkl}. The underlying true galaxy distribution at redshift $z$ is modeled by
\begin{equation}
    p^{\mathrm{true}}_s(z) = z^2\exp{-(z/z_0)^\alpha},
\end{equation}
where $z_0 = 0.3$, $\alpha = 0.9$. The observed galaxy redshift is further blurred by redshift uncertainty $\sigma_z = \sigma_0(1 + z)$ in photometric surveys with $\sigma_0 = 0.05$. 
Assuming that given the true redshift $z$, the observed galaxy redshift $z_{\mathrm{obs}}$ follows a Gaussian distribution
\begin{equation}
    \mathcal{P}(z_{\mathrm{obs}|z}) = \frac{1}{\sqrt{2\pi \sigma^2(z)}}\exp \left[ -\frac{1}{2} \left( \frac{z - z_{\mathrm{obs}}}{\sigma(z)} \right)^2\right].
\end{equation}
The source distribution of galaxies is then
\begin{equation}
    p_s^{i}(z) = p_s^{\mathrm{true}}\int_{z_{\mathrm{min}}^{i}}^{z_{\mathrm{max}}^{i}} dz' \mathcal{P}(z'|z).
\end{equation}
\begin{figure*}[ht!]
    \centering    \includegraphics[width=\textwidth]{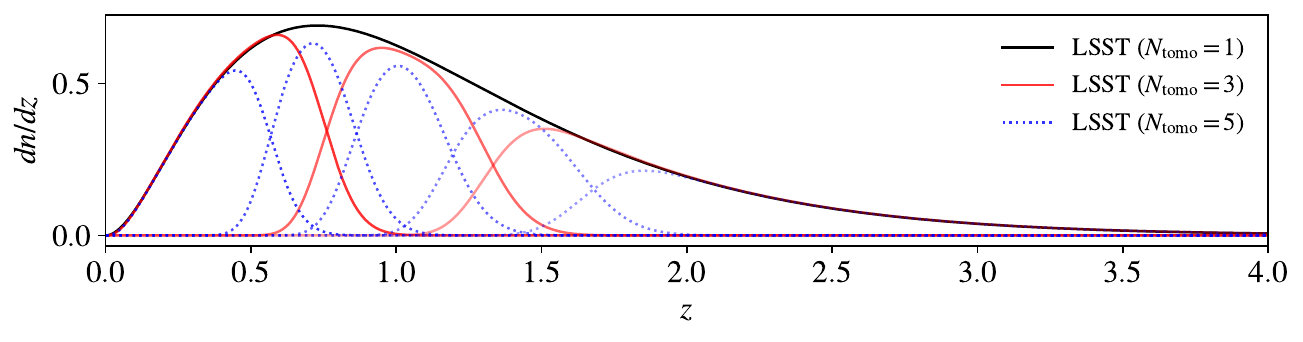}
    \caption{Redshift distribution for different tomographic bin numbers under LSST configuration.}
    \label{fig:Redshift_distribution}
\end{figure*}

Here $z_{\mathrm{min}}$ and $z_{\mathrm{max}}$ is the lower bound and upper bound redshift of each tomographic bin. 
In this work, we investigate how the $f_{\mathrm{NL}}$ constraint changes according to the number of tomographic bins with $N_{\mathrm{tomo}} = 1, 3, 5$. 
For the three tomographic bin case, we divide the galaxies into three bins with equal number density. 
The redshift cut-offs for the three bins are $(0, 0.76), (0.76, 1.34), (1.34,\infty)$. For the five tomographic bin case, the redshift cut-offs are $(0, 0.57), (0.57, 0.87), (0.87, 1.20), (1.20, 1.68), (1.68,\infty)$. 
The redshift distribution of galaxies is normalized in comoving space so that $\int p_s^{i}(\chi)d\chi = 1$. 
With this redshift distribution $p_s^i(\chi)$, the lensing window function $W^{\kappa_g^{(i)}}(\chi)$ for each tomographic bin is then given by Eq.~\eqref{eq:galaxy_lensing_window}. 
We plot the redshift distribution of each tomographic bin in Fig.~\ref{fig:Redshift_distribution}. In this work, we fix the sky coverage at $f_{\mathrm{sky}} = 0.45$.

\subsection{Fisher matrix}
To forecast how $C_{\kappa^g\pi}$ constrains $f_{\mathrm{NL}}$, we perform a Fisher analysis in this section. We assume $f_{\mathrm{NL}}=0$ in the fiducial cosmology and the cross spectrum between lensing convergence fields $\kappa^{g, i}$ and local power spectrum $\pi_k^{\kappa^{g, j}}$ is the only observable, i.e. we do not include cross-correlations with galaxy clustering. 
We first consider an idealized Fisher analysis where $f_{\mathrm{NL}}$ is the only free parameter. 
In Sec.~\ref{sec:marginalization} we then study the impact of marginalizing over the normalization parameter $A_0$, which absorbs theoretical uncertainties in the gravitational non-Gaussianity.

Given a Gaussian likelihood, the Fisher matrix is given by
\begin{equation}
\begin{aligned}
    \mathcal{F} = &\sum_{ijmn}\sum_{kl}\sum_{\ell \in [\ell_{\mathrm{min}}^{\mathrm{soft}}, \ell_{\mathrm{max}}^{\mathrm{soft}}]} \frac{\partial C_{\kappa^{g, i}\pi_k^{\kappa^{g, j}}}(\ell)}{\partial f_{\mathrm{NL}}}  \\ & \times \mathrm{Cov}^{-1}\left[ C_{\kappa^{g, i}\pi_k^{\kappa^{g, j}}}(\ell), C_{\kappa^{g, m}\pi_l^{\kappa^{g, n}}}(\ell) \right] \\ & \times \frac{\partial C_{\kappa^{g, m}\pi_l^{\kappa^{g, n}}}(\ell)}{\partial f_{\mathrm{NL}}}.
\end{aligned}
\end{equation}
Here the partial derivative of cross power spectrum with respect to $f_{\mathrm{NL}}$ is given by Eqs.~(\ref{eq:cl21_final}), (\ref{eq:polynomial_ansatz}), and (\ref{eq:a_fnl}), and the covariance matrix is calculated by Eq.~\eqref{eq:Cov}. The first summation over $i, j, m, n$ runs over all tomographic bins of lensing convergence fields and local power spectrum fields, which varies from $1$ to $N_{\mathrm{tomo}}$. The second summation over $k, l$ accounts for different high-pass filters. The third summation is performed over all multipole moments in the specified range. The uncertainty on $f_{\mathrm{NL}}$ is then given by $\sigma_{f_{\mathrm{NL}}} \geq 1/\sqrt{\mathcal{F}}$. 

For our forecasts, we fix the minimum hard multipole moment of the high-pass filter to $\ell_{\mathrm{min}}^{\mathrm{hard}} = 200$.
This minimum is chosen to ensure the validity of the squeezed-limit approximation. We vary the minimum cutoff of soft multipole moments as $\ell_{\min}^{\mathrm{soft}} = 2, 10, 20$ while fixing the maximum at $\ell_{\mathrm{max}}^{\mathrm{soft}} = 100$. 
Theoretically, most information comes from large scales (small $\ell^{\mathrm{soft}}$) due to the scale-dependent pattern of the $a_{f_{\mathrm{NL}}}$ term (Eq.~\eqref{eq:a_fnl}). 
However, due to survey incompleteness and foreground contamination, the largest scale modes are subject to systematic errors, motivating the consideration of different cutoffs at low $\ell$. 
Regarding the upper bound of $\ell^{\mathrm{soft}}$, the $f_{\mathrm{NL}}$-induced term is negligible on small scales, making it safe to cutoff the cross power spectrum beyond $\ell_{\mathrm{max}}^{\mathrm{soft}} = 100$. 
The width of high pass filter for each $\pi$ field is set to $\Delta \ell = 500$, thus fewer $\pi$ fields are generated, leading to lower computational cost. 
Specifically, The hard multipoles are divided into contiguous high-pass bands with $\{(\ell_{\mathrm{min}}^i), (\ell_{\mathrm{max}}^i)\} = \{(200,700),\ \allowbreak (700,1200),\ \allowbreak (1200,1700),\ \allowbreak (1700,2200),\ \allowbreak (2200,2700),\ \allowbreak (2700,3200),\ \allowbreak (3200,3700),\ \allowbreak (3700,4200), \ \allowbreak  (4200,4700)\}$.
We will show that the width of high-pass filters will not strongly affect the total Fisher information as long as the same scales are included.

\subsection{Fisher Forecast Results}
\begin{figure*}
    \centering
    \includegraphics[width=\textwidth]{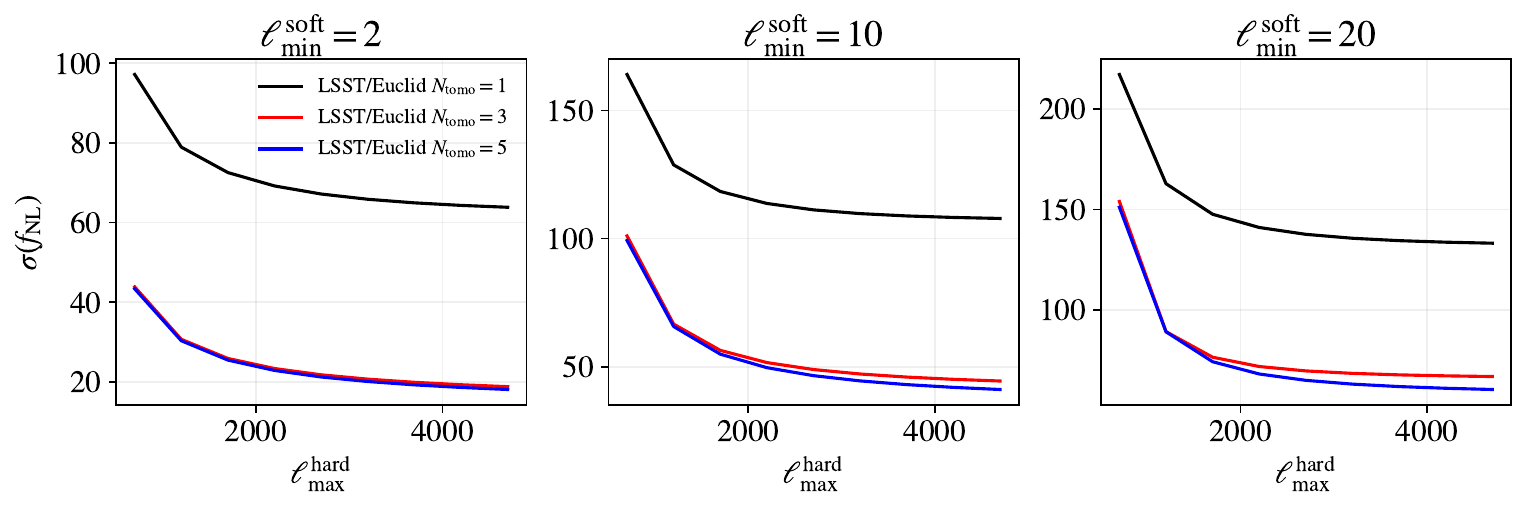}
    \caption{The forecasted constraints on $f_{\mathrm{NL}}$ from cross power spectrum. The curves represent the cumulative information from $\ell_{\mathrm{min}}^{\mathrm{hard}} = 200$ to $\ell_{\mathrm{max}}^{\mathrm{hard}}$. The left panel shows the result with minimum soft mode cut-off $\ell_{\mathrm{min}}^{\mathrm{soft}} = 2$, middle panel with $\ell_{\mathrm{min}}^{\mathrm{soft}} = 10$ and right panel with $\ell_{\mathrm{min}}^{\mathrm{soft}} = 20$. Black curves are result for one tomographic bin. Red curves represent results from three tomographic bins. Blue curves are results for five tomographic bins. 
    }
    \label{fig:Fisher_info}
\end{figure*}

Fig.~\ref{fig:Fisher_info} presents the Fisher forecast results for the LSST configuration, using different numbers of tomographic shear bins. As expected, increasing the number of tomographic bins improves the sensitivity to $f_{\mathrm{NL}}$ by providing additional redshift information. However, the improvement saturates beyond three tomographic bins, consistent with the findings of \citet{2004MNRAS.348..897T, 2024PhRvD.109d3515G}. We also find that the constraint saturates around $\ell_{max} \sim 4000$.

Assuming three tomographic bins and fixing the minimum soft multipole moment to $\ell_{\min}^{\mathrm{soft}} = 10$, we obtain a forecasted precision of $\sigma_{f_\mathrm{NL}} = 44$; for five tomographic bins the result improves to $\sigma_{f_{\mathrm{NL}}} = 41$. Our forecasted sensitivity is somewhat below the bispectrum method of \citep{2024PhRvD.109d3515G} which gives $\sigma_{f_\mathrm{{\mathrm{NL}}}} =16 $ with three tomographic bins and similar choices in shear noise and $\ell$ range. The difference may be due to details in the forecasting choices and a slight sub-optimality of our method compared to the computationally more involved bispectrum analysis.

The expected sensitivity is encouraging. While galaxy clustering from LSST will likely provide a better constraint, weak lensing adds additional signal and an independent consistency check. Furthermore, as we describe below, the method can be combined with other tracers for additional sample variance cancellation.

\subsection{Marginalization of gravitational non-Gaussianity}
\label{sec:marginalization}
In our fiducial bispectrum model (see Eq.~\eqref{eq:bispectrum_model_final}), we introduced a normalization factor \(A_0\) to absorb theoretical uncertainties in the amplitude of $a_0(k, \chi_k)$ (defined in Eq.~\eqref{eq:a_fnl}). These uncertainties mainly originate from baryonic feedback, non-linear evolution, and other small-scale effects that are difficult to model accurately at low redshifts. By treating \(A_0\) as a free parameter during parameter estimation, we effectively marginalize over these residual theoretical systematics and avoid biasing the inferred value of \(f_{\mathrm{NL}}\).

\begin{figure}[htbp]
    \centering
    \begin{minipage}[t]{0.48\textwidth}
        \centering
        \includegraphics[width=0.98\columnwidth]{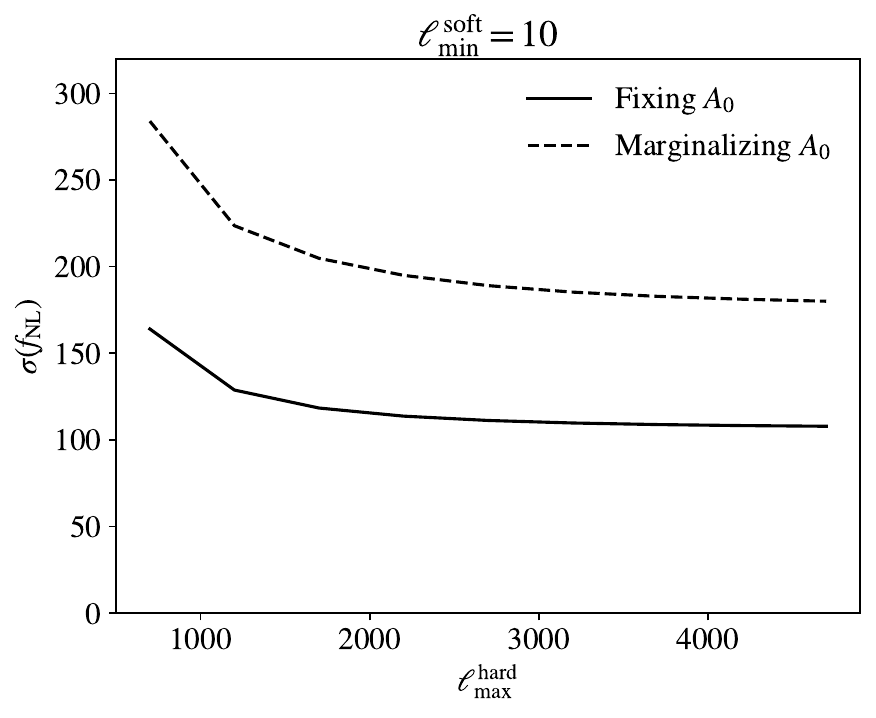}
        \caption{Fisher information for $f_{\mathrm{NL}}$ with and without marginalizing over $A_0$. One single tomographic bin of cosmic shear is assumed. The solid line represents the Fisher information given perfect knowledge of gravitation-induced non-Gaussianity, while the dashed line is the result marginalizes over $A_0$.}
        \label{fig:Fisher_deg_info}
    \end{minipage}
    \begin{minipage}[t]{0.48\textwidth}
        \centering
        \includegraphics[width=0.98\columnwidth]{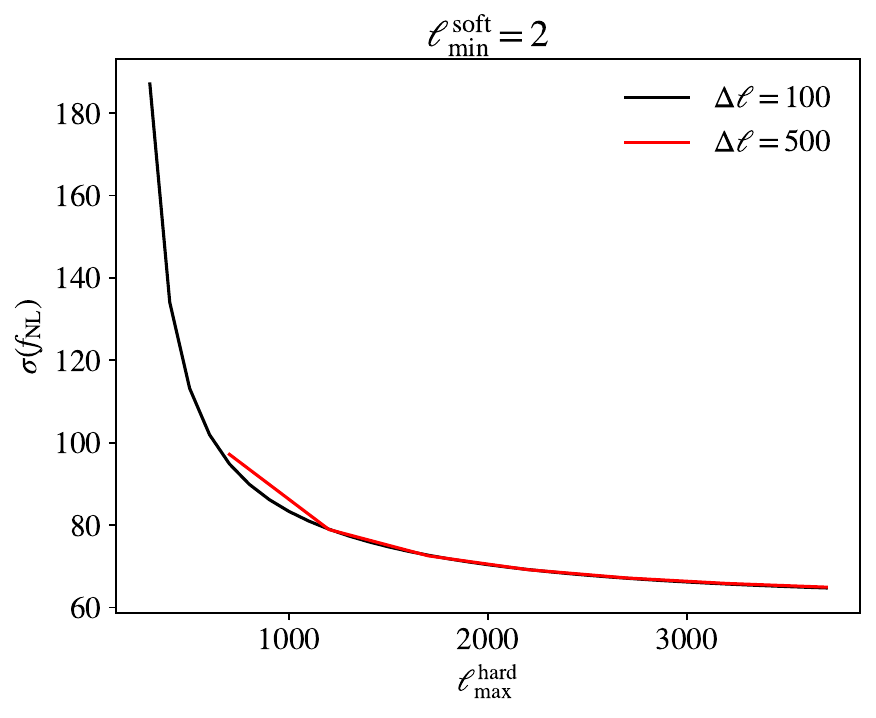}
        \caption{Fisher information for $f_{\mathrm{NL}}$ with high-pass filters of different widths. One single tomographic bin is assumed. The black curve shows the result where each $\pi$ field is generated by a high-pass filter with $\Delta\ell = 100$. The red curve is for $\Delta\ell = 500$.}
        \label{fig:Fisher_diff_info}
    \end{minipage}
\end{figure}

Fig.~\ref{fig:Fisher_deg_info} illustrates the impact of this marginalization. When \(A_0\) is fixed (solid curve), the Fisher information on \(f_{\mathrm{NL}}\) reaches its maximum, corresponding to the ideal case of perfect knowledge of gravitationally induced non-Gaussianity. However, once \(A_0\) is allowed to vary (dashed curve), the degeneracy between \(A_0\) and \(f_{\mathrm{NL}}\) substantially weakens the constraint, reducing the effective Fisher information. This behavior highlights that a portion of the \(f_{\mathrm{NL}}\) signal can be partially mimicked by small-scale amplitude variations encoded in \(A_0\). 

Mitigating this degeneracy, through joint analyses with complementary probes and including auto-powers in addition to cross-powers, or potentially through external priors on \(A_0\) from simulations, will be crucial for achieving precise measurements of primordial non-Gaussianity with future surveys with this method.

\subsection{High-pass Filter with Different Widths}
\label{sec:width}
Fig.~\ref{fig:Fisher_diff_info} shows the forecasted constraint on $f_{\mathrm{NL}}$ for two different choices of the high-pass filter width ($\Delta\ell = 100, 500$). 
In two cases, we include the same total hard-multipoles range but vary the width of each band. Within the same hard-multipoles range, we have five times more $\pi$ fields with $\Delta\ell = 100$ than $\pi$ fields with $\Delta\ell = 500$. 
Despite this difference, the resulting Fisher information is nearly identical across all choices, showing that the estimator is largely insensitive to how the total $\ell$-range is partitioned.

While employing a broader high-pass filter can reduce the complexity of data analysis, an excessively wide filter may fail the squeezed-limit approximation at low $\ell$, thereby introducing modeling inaccuracies. 
Our adopted choice of $\Delta\ell = 500$ provides a good compromise: it preserves the validity of the squeezed-limit approximation while keeping the computational cost reasonable. 

\section{Relation to the large-scale bias model for general $\pi$-fields}
\label{sec:linear_model}

In this work, we developed the $\pi$ field formalism for the special case of the position-dependent power spectrum, following \cite{2023arXiv230503070G}. We calculated the cross-power spectrum of this field using the non-perturbative squeezed limit bispectrum from \citet{2022PhRvD.106l3525G, 2024PhRvD.109d3515G} including the geometric factor from the Gaunt integral Eq. \eqref{eq:pi_multipole}. Our previous work in \cite{2023arXiv230503070G} took a somewhat different approach, by postulating and validating a bias model that should be valid for any field which is sensitive to local primordial power on large scales. We worked with the more explicit calculation to clarify the role of the geometric projection to the sphere and of light-cone integration of Gaussian and non-Gaussian biases in weak lensing. However, the general $\pi$-field bias model has the advantage that it generalizes straight forward to other fields, in particular the neural network approach of \cite{2023PhRvD.107f1301G}, which we aim to apply to weak lensing in future work. In this section, we compare the two approaches.

\subsection{$\pi$ fields in 3D Cartesian frame}

We briefly recall the arguments behind the $\pi$-field bias model. In a universe with non-zero local PNG, the primordial potential field $\Phi(\bm{x})$ is defined by Eq.~\eqref{eq:fnl_def}. 
The quadratic term couples Fourier modes, producing a non-zero bispectrum and higher-point statistics. 
This leads to a large-scale modulation of small-scale power.
To demonstrate this, we decompose the Gaussian potential into long- and short-wavelength parts,
\begin{equation}
    \Phi_G(\bm{x}) = \Phi_l(\bm{x}) + \Phi_s(\bm{x}).
\end{equation}
Inserting this into Eq.~\eqref{eq:fnl_def} gives (\cite{smith2012halo})
\begin{equation}
\begin{aligned}
    \Phi(\bm{x}) = &
    \underbrace{\Phi_l(\bm{x}) + f_{\mathrm{NL}}\!\left[\Phi_l(\bm{x})^2 - \langle \Phi_l^2\rangle\right]}_{\text{long}} \; \\ & +\;
    \underbrace{\left[1 + 2 f_{\mathrm{NL}} \Phi_l(\bm{x})\right]\Phi_s(\bm{x}) 
    + f_{\mathrm{NL}}\!\left[\Phi_s(\bm{x})^2 - \langle \Phi_s^2 \rangle \right]}_{\text{short}}.
\end{aligned}
\end{equation}
The term $\left[1 + 2 f_{\mathrm{NL}}\Phi_l(\bm{x})\right]\Phi_s(\bm{x})$ gives rise to the characteristic large-scale modulation of small-scale modes of local non-Gaussianity. We can interpret this term as a spatially varying amplitude of matter fluctuations, 

\begin{equation}
\begin{aligned}
    \Delta\sigma^{\mathrm{loc}}_8(\bm{x}) &= \sigma^{\mathrm{loc}}_8(\bm{x}) - \bar{\sigma}_8\\
    &= \bar{\sigma}_8 2 f_{\mathrm{NL}} \Phi_l(\mathbf{x}) 
\end{aligned}
\end{equation}
On large scales, any field field $\pi(\bm{x})$ that is sensitive to both the long-wavelength matter fluctuations and to the amplitude of matter perturbations $\sigma_8$ can be described by the local bias expansion to first order in the perturbations as
\begin{equation}
\begin{aligned}
    \pi(\bm{x}) 
    & = \bar{\pi} + \frac{\partial \pi}{\partial \delta_l}\,\delta_l(\bm{x})
    + \frac{\partial \pi}{\partial \sigma_8}\,\Delta\sigma_8(\bm{x})\\
    &=  \bar{\pi} + \frac{\partial \pi}{\partial \delta_l}\,\delta_l(\bm{x})
    + 2 \bar{\pi}f_{\mathrm{NL}}\frac{\partial \log \pi}{\partial \log \sigma_8}\, \Phi_l(\bm{x}) 
\end{aligned}
\end{equation}
Introducing the ``Gaussian bias" $b_\pi \equiv \partial \pi/\partial \delta_l$ and the ``non-Gaussian bias" \(\beta_\pi \equiv 2\bar{\pi} \partial \ln \pi / \partial \ln \sigma_8\), the above equation becomes
\begin{equation}
    \pi(\bm{x}) = \bar{\pi} + b_\pi\, \delta_l(\bm{x}) + \beta_\pi\, f_{\mathrm{NL}}\, \Phi_l(\bm{x}).
\end{equation}

Going to Fourier space and using Poisson’s equation 
\(\delta_l(\bm{k},\chi) = \alpha(k,\chi)\,\Phi_l(\bm{k})\) with 
\(\alpha(k,\chi) = 2 k^2 T(k) D(\chi)/(3\Omega_m H_0^2)\), we obtain
\begin{equation}
    \pi(\bm{k},\chi) 
    = \left[\, b_\pi + \,\beta_\pi\,\frac{f_{\mathrm{NL}}}{\alpha(k,\chi)} \right]\delta_l(\bm{k},\chi).
    \label{eq:model}
\end{equation}
The second term in the parenthesis is the characteristic signature of local PNG, which scales as $k^{-2}$ on large scales, making it distinct from any linear or non-linear Gaussian bias terms. Recent work related to the $\pi$-field approach also includes \cite{Chen:2024exy,2025OJAp....8E...6H}.

\subsection{Adapting $\pi$ fields to spherical coordinates}
To apply this method to wide-field photometric surveys (both clustering and weak lensing), we adapt the model from a 3D Cartesian framework to the binned spherical coordinates. Expanding $\pi$ in spherical Fourier–Bessel modes and using Eq.~\eqref{eq:projected_map}, we have
\begin{equation}
\begin{aligned}
    \pi_{\ell m} 
    = &  4\pi i^\ell \!\int_0^\infty\! d\chi\, W^\delta(\chi)\!\int \frac{d^3\bm{k}}{(2\pi)^3}\, 
       j_\ell(k\chi)\, Y_{\ell m}^\ast(\hat{\bm{k}})\,\pi(\bm{k},\chi) \label{eq:expansion}\\
    = & 4\pi i^\ell \!\int_0^\infty\! d\chi\, W^\delta(\chi)\!\int \frac{d^3\bm{k}}{(2\pi)^3}\, 
       j_\ell(k\chi)\, Y_{\ell m}^\ast(\hat{\bm{k}})\, 
        \\ & \times \Big[ b_\pi + \,\beta_\pi\,\frac{f_{\mathrm{NL}}}{\alpha(k,\chi)} \Big]\delta(\bm{k},\chi),
\end{aligned}
\end{equation}
where $Y_{\ell m}$ are spherical harmonics and $j_\ell$ are spherical Bessel functions. The first step is the spherical harmonics expansion of $\pi(\bm{x})$ in Fourier space. 
In the second step we applied Eq.~\eqref{eq:model}. 
This yields the large-scale prediction for the angular cross-power spectrum between the projected field $\delta(\bm{n})$ and the local power spectrum field $\pi(\bm{n})$:
\begin{equation}
\begin{aligned}
    C_{\delta\pi}(\ell) 
    = &  \frac{2}{\pi}\int k^2 dk \int d\chi_1\, W^\delta(\chi_1) j_\ell(k\chi_1)
       \\ & \times \int d\chi_2\, W^\delta(\chi_2) j_\ell(k\chi_2)
       \\ & \times \left[ b_\pi + \,\beta_\pi\,\frac{f_{\mathrm{NL}}}{\alpha(k,\chi_1)} \right]
      P(k,\chi_1,\chi_2).
    \label{eq:simplifiedlinear}
\end{aligned}
\end{equation}
Both $b_\pi$ and $\beta_\pi$ are  redshift dependent and scale dependent, and for wide projection kernels, such as in weak lensing, different values will contribute. 

Furthermore, in weak lensing there is always a contribution to the kernel from lower redshifts, where the probed $k$ are too small-scale to be entirely modeled with this ``linear+$f_{\mathrm{NL}}$" bias model. Fortunately, as we have seen above in our simulations, a two parameter model with a fitted linear bias amplitude $A_0$ and an $f_{\mathrm{NL}}$ amplitude does fit the simulated weak lensing power spectra to good approximation on large scales. 

For large scales, we approximate the Gaussian and non-Gaussian biases by effective (redshift averaged) constants \(\hat{b}_\pi\) and \(\hat{\beta}_\pi\) which can be calibrated with simulations, leading to
\begin{equation}
    C_{\delta\pi}(\ell) 
    = \hat{b}_\pi\, C_{\delta\delta}(\ell) 
    + \hat{\beta}_\pi\, f_{\mathrm{NL}}\, C_{\delta\delta}^{l\text{--}s}(\ell),
    \label{eq:linear_model}
\end{equation}
with
\begin{equation}
\begin{aligned}
    C_{\delta\delta}^{l\text{--}s}(\ell) 
    = &  \frac{2}{\pi}\int k^2 dk \int d\chi_1\, W^\delta(\chi_1) j_\ell(k\chi_1)
       \\ & \times \int d\chi_2\, W^\delta(\chi_2) j_\ell(k\chi_2)\,
      \frac{P(k,\chi_1,\chi_2)}{\alpha(k,\chi_1)}.
    \label{eq:l-s}
\end{aligned}
\end{equation}
Eq.~\eqref{eq:linear_model} successfully captures the leading large-scale behavior. 
The first term $\hat{b}_\pi C_{\delta \delta}(\ell)$ refers to the angular cross power spectrum in the fiducial cosmology ($f_{\mathrm{NL}} = 0$). 
The second term $\hat{\beta}_\pi f_{\mathrm{NL}} C_{\delta \delta}^{l-s}(\ell)$ arises from non-zero $f_{\mathrm{NL}}$. 

\subsection{Relation to the non-perturbative model with explicit Gaunt factor projection}

Comparing the linear model Eq.~\eqref{eq:simplifiedlinear} and the non-perturbative result Eq.~\eqref{eq:cl21_final}, we see that in the latter we explicitly evaluate the geometric projection due to the squaring of the field using the Gaunt factor. Note that we are not squaring the local matter field and then integrating over that squared field, but rather we are squaring the matter field after integrating it (with the lensing kernel). In addition to explicitly accounting for geometry, the non-perturbative model also allows us to model the redshift dependence of biases in an analytic way (up to an overall amplitude of the Gaussian bias $A_0$), using Eq. \eqref{eq:a_0_grav} and \eqref{eq:a_fnl}.

\begin{figure*}[ht!]
    \centering    \includegraphics[width=\textwidth]{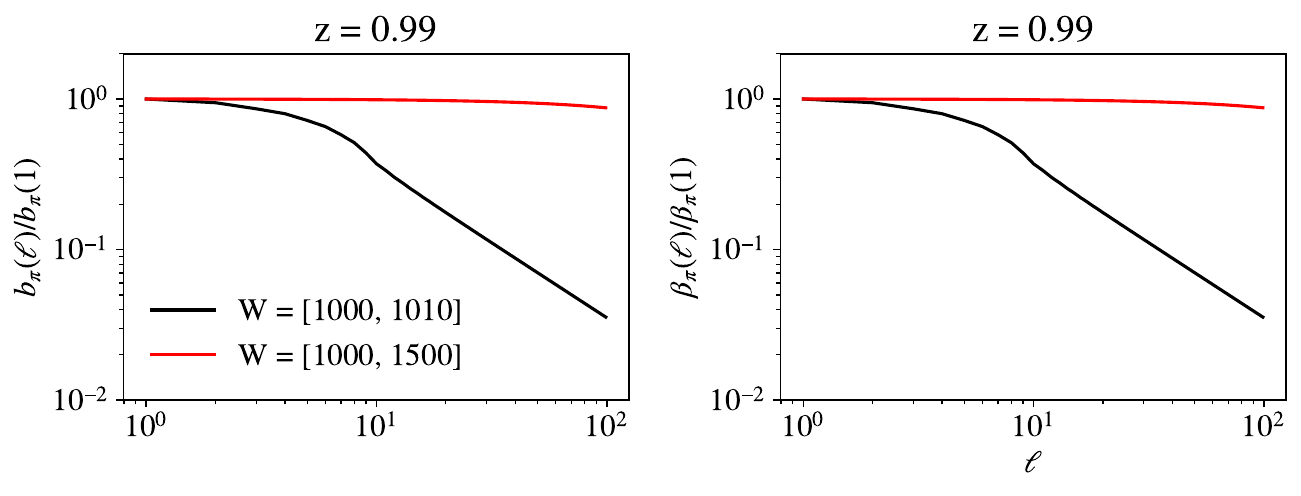}
    \caption{Gaussian bias (left) and non-Gaussian bias (right) for the $\pi$ field for a matter shell.}
    \label{fig:Bias_figure}
\end{figure*}

We first compare the two models in the thin matter-shell limit. For a thin shell with $W^\delta(\chi)=\delta(\chi-\chi_0)$, the effective Gaussian and non-Gaussian “bias functions’’ inferred from the non-perturbative formula are
\begin{equation}
\begin{aligned}
    {b}_\pi(\ell) \;=&\; \frac{1}{\chi_0^2}
    \sum_{\ell_1\ell_2}\frac{(2\ell_1\!+\!1)(2\ell_2\!+\!1)}{4\pi}
    \begin{pmatrix} \ell_1 & \ell_2 & \ell \\ 0 & 0 & 0 \end{pmatrix}^{\!2} \\ &  \times
    W^{\mathrm{HP}}(\ell_1) W^{\mathrm{HP}}(\ell_2)\;
    a_0(k,\chi_0)\, P(k,\chi_0), 
    \label{eq:b_G}
\end{aligned}
\end{equation}
\begin{equation}
\begin{aligned}
    {\beta}_\pi(\ell) \;=&\; \frac{1}{\chi_0^2}
    \sum_{\ell_1\ell_2}\frac{(2\ell_1\!+\!1)(2\ell_2\!+\!1)}{\pi}
    \begin{pmatrix} \ell_1 & \ell_2 & \ell \\ 0 & 0 & 0 \end{pmatrix}^{\!2} \\ & \times 
    W^{\mathrm{HP}}(\ell_1) W^{\mathrm{HP}}(\ell_2)\;
    \frac{\partial\!\ln P(k,\chi_0)}{\partial\!\ln \sigma_8^2}\, P(k,\chi_0).
    \label{eq:b_NG}
\end{aligned}
\end{equation}
Fig.~\ref{fig:Bias_figure} shows $b_\pi(\ell)$ and $\beta_\pi(\ell)$ at $z=0.47$ and $z=1.14$ for two high-pass bands 
$(\ell_{\min}^{\mathrm{hard}},\ell_{\max}^{\mathrm{hard}})=\{(1000,1010),(1000,1500)\}$. In Appendix~\ref{sec:Asymptotic}, we provide an  analytic analysis of the squeezed limit behavior of the geometric projection factor, finding that it goes as $1/\ell$ for a delta function high-pass filter. On the other hand, as shown in Fig. \ref{fig:Bias_figure}, the geometric factor softens progressively with a wider high-pass filter, and is approximately constant in the case of $\ell=(1000,1500)$, approaching the constant bias of the linear model in Eq. \eqref{eq:linear_model}.

\begin{figure*}[ht!]
    \centering    \includegraphics[width=\textwidth]{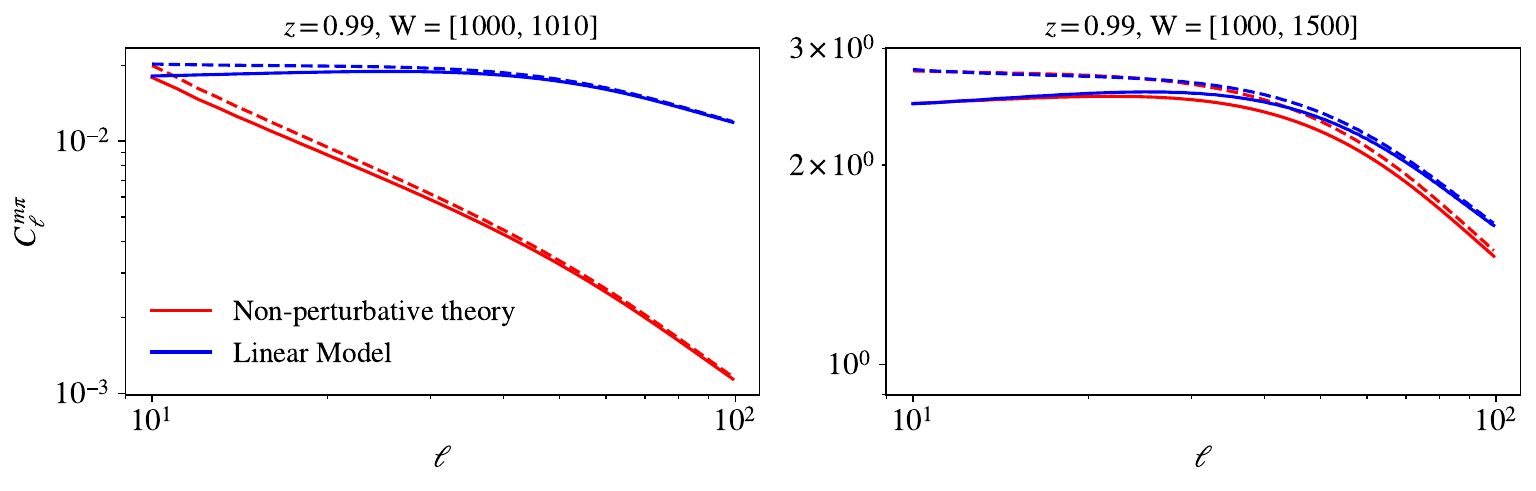}
    \caption{The comparison of two models (red for non-perturbative model and blue for linear model) at redshift $z = 0.99$. The solid lines are cosmology with $f_{\mathrm{NL}} = 0$, dashed lines for $f_{\mathrm{NL}} = 100$. The left panel is the results for narrow high-pass filter $(\ell_\mathrm{min}, \ell_\mathrm{max}) = ([1000, 1010])$. The right panel is for a wide high-pass filter $(\ell_\mathrm{min}, \ell_\mathrm{max}) = ([1000, 1500])$. The agreement is good for a wide high-pass filter, but breaks down for a narrow high-pass filter where the geometric projection factor is steep.}
    \label{fig:Compare_m1}
\end{figure*}

\begin{figure*}[ht!]
    \centering    \includegraphics[width=\textwidth]{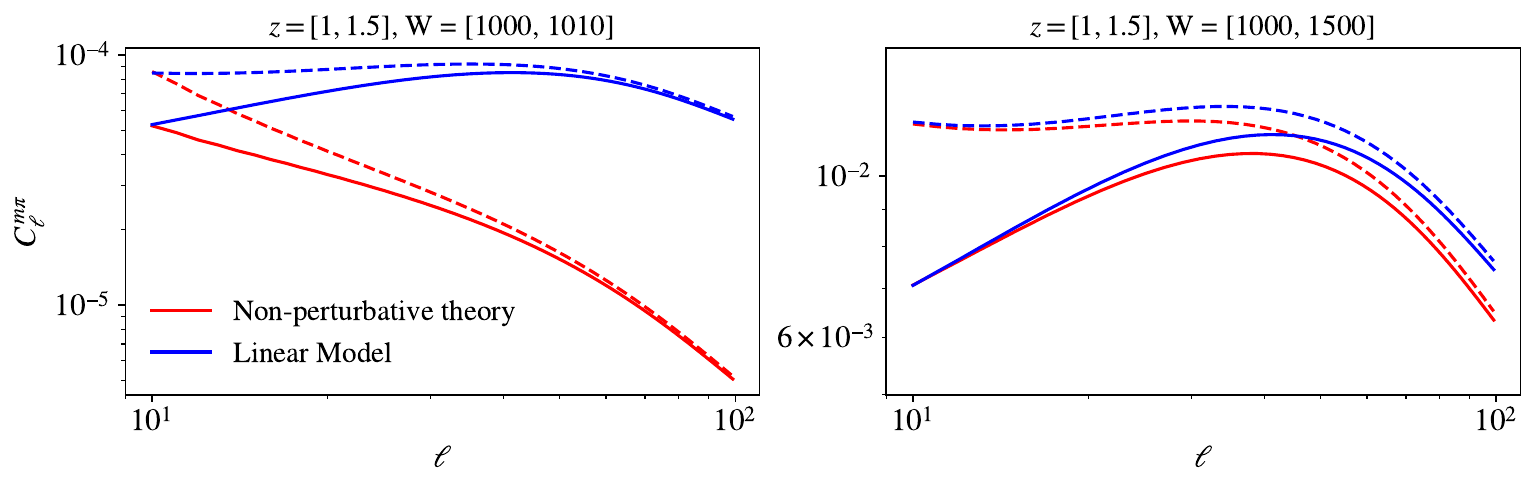}
    \caption{The comparison of two models (red for non-perturbative model and blue for linear model) for the matter field averaged between $z = 0.99$ to $z = 1.5$. The solid lines are cosmology with $f_{\mathrm{NL}} = 0$, dashed lines for $f_{\mathrm{NL}} = 100$. The left panel is the results for narrow high-pass filter $(\ell_\mathrm{min}, \ell_\mathrm{max}) = ([1000, 1010])$. The right panel is for a wide high-pass filter $(\ell_\mathrm{min}, \ell_\mathrm{max}) = ([1000, 1500])$.}
    \label{fig:Compare_m2}
\end{figure*}

\begin{figure*}[ht!]
    \centering    \includegraphics[width=\textwidth]{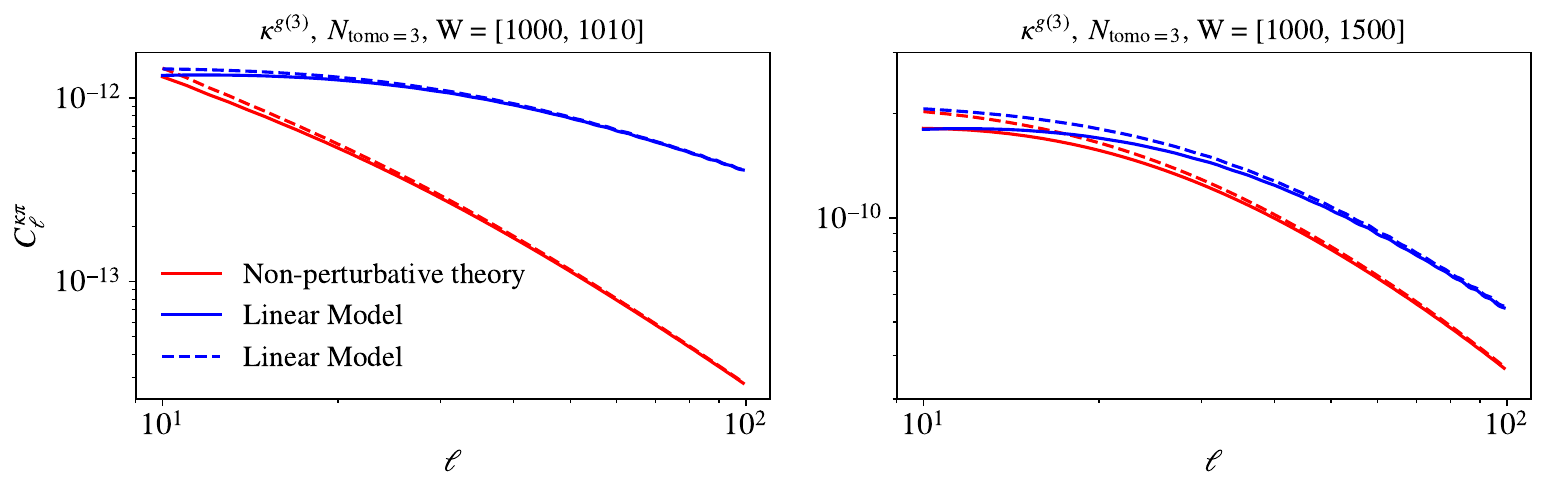}
    \caption{The comparison of two models (red for non-perturbative model and blue for linear model) for the third lensing convergence field with $N_{\mathrm{tomo}} = 3$. The solid lines are cosmology with $f_{\mathrm{NL}} = 0$, dashed lines for $f_{\mathrm{NL}} = 100$. The left panel is the results for narrow high-pass filter $(\ell_\mathrm{min}, \ell_\mathrm{max}) = ([1000, 1010])$. The right panel is for a wide high-pass filter $(\ell_\mathrm{min}, \ell_\mathrm{max}) = ([1000, 1500])$.}
    \label{fig:Compare_k}
\end{figure*}

In Figs.~\ref{fig:Compare_m1}, \ref{fig:Compare_m2}, and \ref{fig:Compare_k}, we compare the linear model with the non-perturbative result for several cases, including matter shells of different widths and the lensing convergence field. 
The solid curves correspond to the $f_{\mathrm{NL}} = 0$ cosmology, while the dashed curves show the $f_{\mathrm{NL}} = 100$ case. 
The effective biases $b_\pi$ and $\beta_\pi$ are fitted to reproduce the large-scale behavior of the non-perturbative model. 
As is shown, the linear model approaches the non-perturbative prediction as the high-pass filter becomes wider. In summary, the $\pi$-field model of Eq.~\eqref{eq:simplifiedlinear} is correct on large scales, and could be used for example for a neural network-based $\pi$ field as in \cite{2023PhRvD.107f1301G}, but the range of validity needs to be carefully evaluated. For example, it could be useful to analyze differential redshift bins, to make the $\pi$ field more local in radial direction, and improve the model fit. We defer this to future work.

\section{Discussion}
\label{sec:discussion}

In this work we have developed a novel $f_{\mathrm{NL}}$ estimator for weak lensing maps. The method is based on the local power spectrum field $\pi(\bm{n})$ recently proposed by \citet{2023arXiv230503070G}. We extended this statistics from 3D Cartesian boxes to binned spherical coordinates, derived its angular cross-power spectrum with projected large-scale structure fields, and built a non-perturbative model for that cross-spectrum in terms of the squeezed-limit matter bispectrum.

Unlike other approaches that rely on higher-order bispectrum statistics, our analysis is based on the power spectrum, significantly reducing computational complexity, while still being optimal in the squeezed limit.
This simplicity allows for a robust estimation of $f_{\mathrm{NL}}$ and makes our method a powerful addition to, and cross-check of, the traditional halo field analysis. We have successfully verified our model with simulated matter fields at low redshift, showing excellent agreement with the theoretical predictions.

Using this model, we forecasted the $f_{\mathrm{NL}}$ sensitivity for an LSST-like weak lensing survey. For five tomographic source bins and $\ell_{\mathrm{min}} = 10$, we forecast a precision of $\sigma_{f_{\mathrm{NL}}} \simeq 41$. For a more aggressive multipole cut with $\ell_{\mathrm{min}} = 2$, our method achieves a constraint of $\sigma_{f_{\mathrm{NL}}} = 18$ when using $N_{\mathrm{tomo}} = 5$ tomographic bins.

The method we described here is a special case of the more general $\pi$-field methodology for squeezed limit non-Gaussianity measurements \citep{2023PhRvD.107f1301G, 2023arXiv230503070G, 2szy-wypg}. Our ultimate goal is to combine several different large-scale fields in a combined likelihood to obtain the tightest possible $f_{NL}$ constraints. This likelihood would contain galaxy clustering $\delta_g$, weak lensing $\kappa$, the CMB lensing potential $\phi$ and the kSZ velocity reconstruction field $v^{kSZ}$. Both the lensing fields and the kSZ velocity fields are sensitive to $f_{NL}$ via sample variance cancellation (see  \cite{Seljak:2008xr,Schmittfull:2017ffw,2023PhRvD.108h3522M,Munchmeyer:2018eey}. Further, as we have described, both galaxy clustering and lensing can be used to construct $\pi$ fields which are sensitive to the large-scale modulation of small-scale power induced by local non-Gaussianity. The combination of several probes also helps to break parameter degeneracies between biases. The present work adds a new element to this general program. While we focused here on local non-Gaussianity, our method also extends to other squeezed limit scaling behaviors (see e.g. \cite{Green:2023uyz}).

In summary, the cross-spectrum between a projected field and its associated local power spectrum field, $C_{\delta \pi}(\ell)$, provides a practical probe of primordial non-Gaussianity. It captures the squeezed-limit bispectrum information that is sensitive to $f_{\mathrm{NL}}$, while retaining the analysis simplicity of a two-point statistic. This makes it a promising path toward competitive, survey-ready constraints on physics of the primordial universe.

\begin{acknowledgments}
We thank Utkarsh Giri for his advise and initial collaboration, and Kendrick Smith and Samuel Goldstein for helpful discussions during the development of this work. M.M. and S.Z. acknowledge the support by the U.S. Department of Energy, Office of Science, Office of High Energy Physics under Award Number DE-SC0017647, the support by the National Science Foundation (NSF) under Grant Number 2307109 and 2509873 and
the Wisconsin Alumni Research Foundation (WARF). 
\end{acknowledgments}

\appendix

\section{Gaunt Integral Simplification}
\label{sec:Gaunt}
The Gaunt integral $\mathcal{G}_{m_1m_2m_3}^{l_1l_2l_3}$ can be written as 
\begin{equation}
\begin{aligned}
    \mathcal{G}_{m_1m_2m_3}^{\ell_1\ell_2\ell_3} = &  \sqrt{\frac{(2\ell_1 + 1)(2\ell_2 + 1)(2\ell_3 + 1)}{4\pi}} \\ & \times \begin{pmatrix} \ell_1 & \ell_2 & \ell_3\\ 0 & 0 & 0\end{pmatrix}\begin{pmatrix} \ell_1 & \ell_2 & \ell_3\\ m_1 & m_2 & m_3\end{pmatrix},
\end{aligned}
\end{equation}
here the bracket matrix refers to Wigner 3-j symbol. For the Wigner 3-j symbol, we have the following property with which we can simplify the sum of the square of Gaunt integrals.
\begin{equation}
\begin{aligned}
    (\mathcal{G}_{m_1m_2m}^{\ell_1\ell_2\ell})^2 =& (2\ell + 1)\delta_{m, m'}\{ \ell_1 \ \ell_2 \ \ell \} \\ & \times \sum_{m_1m_2}\begin{pmatrix} \ell_1 & \ell_2 & \ell \\ m_1 & m_2 & m\end{pmatrix}\begin{pmatrix} \ell_1 & \ell_2 & \ell'\\ m_1 & m_2 & m'\end{pmatrix}\delta_{\ell, \ell'},
\end{aligned}
\end{equation}
here $\{ \ell_1 \ \ell_2 \ \ell\}$ is one if they satisfy triangle conditions, otherwise zero. Then
\begin{equation}
\begin{aligned}
    \sum_{\ell_1\ell_2m_1m_2}(\mathcal{G}_{m_1m_2m}^{\ell_1\ell_2\ell})^2 F(\ell_1, \ell_2, \ell) 
    = & \sum_{\ell_1 \ell_2}F(\ell_1, \ell_2, \ell) \\ & \times\frac{(2\ell_1 + 1)(2\ell_2 + 1)}{4\pi} \\ & \times \begin{pmatrix} \ell_1 & \ell_2 & \ell\\ 0 & 0 & 0\end{pmatrix}^2
\end{aligned}
\end{equation}

\vspace{0.5em}
\section{Tiling Effect in the Ulagam simulations}
\label{sec:Tiling}

\begin{figure*}[ht!]
    \centering    \includegraphics[width=\textwidth]{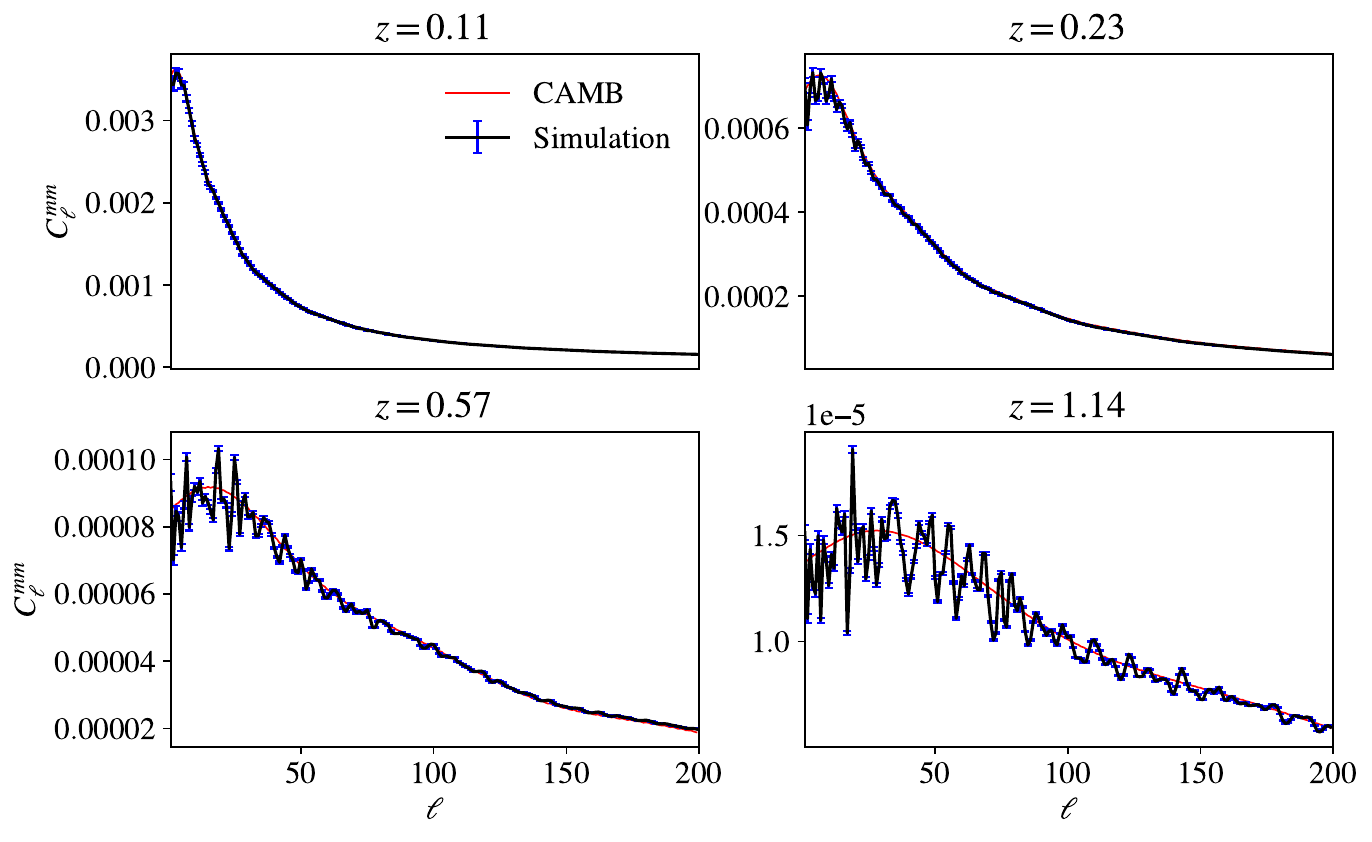}
    \caption{The matter power spectrum at multiple redshifts. The black lines are results averaged over 1000 simulations, red lines are calculated from {\tt CAMB} with Eq.~\eqref{eq:Power_Accu}, blue bars are 1-$\sigma$ uncertainties. The apparent oscillations are not due to cosmic variance but due to the simulation tiling.}
    \label{fig:patch}
\end{figure*}

\begin{figure*}[ht!]
    \centering    \includegraphics[width=\textwidth]{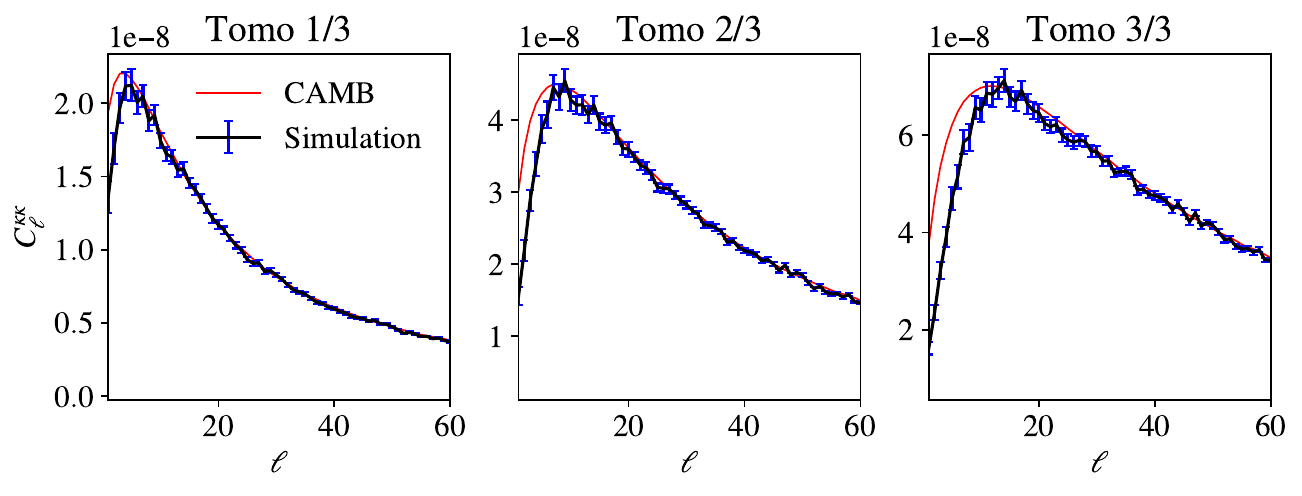}
    \caption{The power spectrum of lensing fields. The black lines are results averaged over 100 simulations, red lines are calculated from {\tt CAMB} with Eq.~\eqref{eq:Power_Accu}, blue bars are 1-$\sigma$ uncertainties.}
    \label{fig:patch_lensing}
\end{figure*}

In Fig.~\ref{fig:patch}, we plot the matter power spectrum at different redshifts both from simulations and theoretical calculation. As the figure shows, the tiling results in the oscillating features of the power spectrum on large scales. At low redshift, most of the particles are from one single simulation box, thus not affected by tiling. In comparison, the matter power spectrum at higher redshift suffers from tiling severely. Fig.~\ref{fig:patch_lensing} shows the lensing power spectrum for different tomographic bins. Generally, the simulation result underestimates the power spectrum due to tiling. For the first tomographic bin (lowest redshift), only modes on largest scales are affected. But for the tomographic bins at higher redshift this effect is no longer negligible. 

This is not a shortcoming of the Ulagam simulations, which were not constructed for scale-dependent bias by their authors, but rather a natural consequence of tiling. We illustrate the effect here to show why a precision test of our method at higher redshift cannot be performed with this simulation set.

\vspace{0.5em}

\section{Asymptotic behavior of the geometric factor}
\label{sec:Asymptotic}

When the high-pass filter $W_{\mathrm{HP}}
(\ell)$ is extremely narrow, the geometric factor could be further simplified. 
In this section we show the details of this asymptotic behavior. We include this analytic analysis to clarify how the geometric coupling alone scales in this limit and to understand why narrow high-pass filters can produce departures from the simplified large-scale model.

Suppose the high-pass filter selects one single spherical harmonic $\ell_{\mathrm{HP}}$, that is
\begin{equation}
    W^{\mathrm{HP}}(\ell) = \begin{cases}
    1, & \text{if } \ \ell = \ell_{\mathrm{HP}}  \\
    0, & \text{otherwise}.
\end{cases}
\end{equation}
Then the geometric factor $\mathcal{N}$ becomes
\begin{equation}
\begin{aligned}
    \mathcal{N} = &  W^{\mathrm{HP}}(\ell_1)W^{\mathrm{HP}}(\ell_2)\sum_{\ell_1\ell_2}(2\ell_1 + 1)(2\ell_2 + 1)\begin{pmatrix} \ell_1 & \ell_2 & \ell\\ 0 & 0 & 0\end{pmatrix}^2 \\  = &  (2\ell_{\mathrm{HP}} + 1)^2\begin{pmatrix} \ell_1 & \ell_2 & \ell\\ 0 & 0 & 0\end{pmatrix}^2
\end{aligned}
\end{equation}
In this work, generally the multipole $\ell$ of the cross-powers are much smaller than the multipoles filtered out by the high-pass filter, $\ell \ll \ell_{\mathrm{HP}}$. 
Under this assumption we can apply the following approximation \citep{1988qtam.book.....V}:
\begin{equation}
    \begin{pmatrix} \ell_1 & \ell_2 & \ell_3\\ m_1 & m_2 & m_3\end{pmatrix} \approx (-1)^{\ell_3 + m_3}\frac{d_{m_1, \ell_3 - \ell_2}^{\ell_1}(\theta)}{\sqrt{2\ell_3 + 1}}\, \ ( \ell_1 \ll \ell_2, \ell_3).
\end{equation}
Here $\cos(\theta) = -2m_3 / (2\ell_3 + 1)$ equals zero in our scenario, thus $\theta = \frac{\pi}{2}$. 
$d_{m, m'}^\ell(\ell)$ is the d-matrix. Thus we have
\begin{equation}
\begin{aligned}
    \mathcal{N} \approx &  (2\ell_{\mathrm{HP}} + 1) d_{0, 0} ^ 0(\frac{\pi}{2}) \\ = & (2\ell_{\mathrm{HP}} + 1)P_{\ell}^0(0)^2  \\ \approx & (2\ell_{\mathrm{HP}} + 1)\frac{2}{\pi\ell}\cos\frac{\ell \pi}{2} \propto \frac{1}{\ell}\quad \text{when} \ \ell \gg 1.
\end{aligned}
\end{equation}

This $1/\ell$ factor is a purely geometric effect that does not appear in the simplified large-scale model, which assumes both the linear bias and non-linear bias as constant in the large scale limit.

\begin{figure}[ht!]
    \centering    \includegraphics[width=\columnwidth]{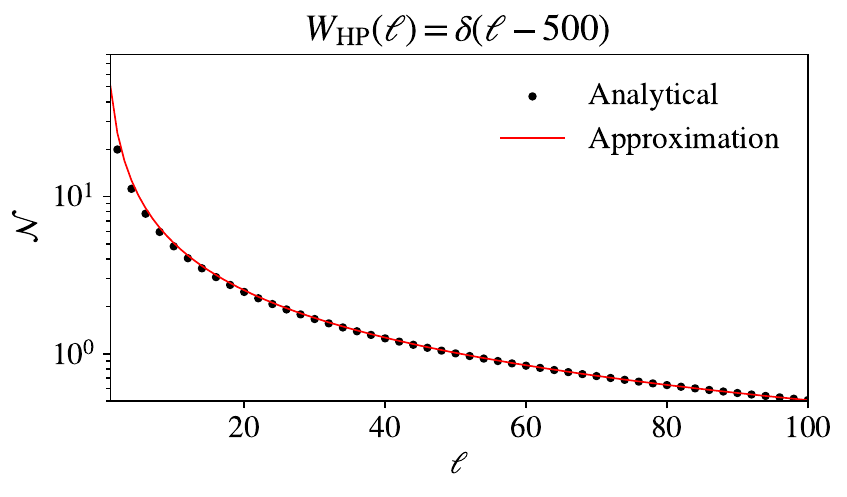}
    \caption{The geometric factor $\mathcal{N}$ calculated in both analytical and approximated ways.}
    \label{fig:Geo_narrow_approx}
\end{figure}
\vspace{0.5em}
\bibliography{apssamp}

@ARTICLE{2023PhRvD.108h3522M,
       author = {{McCarthy}, Fiona and {Madhavacheril}, Mathew S. and {Maniyar}, Abhishek S.},
        title = "{Constraints on primordial non-Gaussianity from halo bias measured through CMB lensing cross-correlations}",
      journal = {\prd},
         year = 2023,
        month = oct,
       volume = {108},
       number = {8},
          eid = {083522},
        pages = {083522},
          doi = {10.1103/PhysRevD.108.083522},
archivePrefix = {arXiv},
       eprint = {2210.01049},
 primaryClass = {astro-ph.CO},
       adsurl = {https://ui.adsabs.harvard.edu/abs/2023PhRvD.108h3522M}
}

@article{Esposito:2019jkb,
    author = "Esposito, Angelo and Hui, Lam and Scoccimarro, Roman",
    title = "{Nonperturbative test of consistency relations and their violation}",
    eprint = "1905.11423",
    archivePrefix = "arXiv",
    primaryClass = "astro-ph.CO",
    doi = "10.1103/PhysRevD.100.043536",
    journal = "Phys. Rev. D",
    volume = "100",
    number = "4",
    pages = "043536",
    year = "2019"
}

@ARTICLE{Peloso:2013zw,
       author = {{Peloso}, Marco and {Pietroni}, Massimo},
        title = "{Galilean invariance and the consistency relation for the nonlinear squeezed bispectrum of large scale structure}",
      journal = {J. Cosmol. Astropart. Phys.},
         year = 2013,
        month = may,
       volume = {2013},
       number = {5},
          eid = {031},
        pages = {031},
          doi = {10.1088/1475-7516/2013/05/031},
archivePrefix = {arXiv},
       eprint = {1302.0223},
 primaryClass = {astro-ph.CO},
       adsurl = {https://ui.adsabs.harvard.edu/abs/2013JCAP...05..031P}
}

@ARTICLE{2024JCAP...03..062A,
       author = {{Anbajagane}, Dhayaa and {Chang}, Chihway and {Lee}, Hayden and {Gatti}, Marco},
        title = "{Primordial non-Gaussianities with weak lensing: information on non-linear scales in the ULAGAM full-sky simulations}",
      journal = {J. Cosmol. Astropart. Phys.},
         year = 2024,
        month = mar,
       volume = {2024},
       number = {3},
          eid = {062},
        pages = {062},
          doi = {10.1088/1475-7516/2024/03/062},
archivePrefix = {arXiv},
       eprint = {2310.02349},
 primaryClass = {astro-ph.CO},
       adsurl = {https://ui.adsabs.harvard.edu/abs/2024JCAP...03..062A}
}

@ARTICLE{2023arXiv230503070G,
       author = {{Giri}, Utkarsh and {M{\"u}nchmeyer}, Moritz and {Smith}, Kendrick M.},
        title = "{Constraining $f_{NL}$ using the Large-Scale Modulation of Small-Scale Statistics}",
      journal = {arXiv e-prints},
         year = 2023,
        month = may,
          eid = {arXiv:2305.03070},
        pages = {arXiv:2305.03070},
          doi = {10.48550/arXiv.2305.03070},
archivePrefix = {arXiv},
       eprint = {2305.03070},
 primaryClass = {astro-ph.CO},
       adsurl = {https://ui.adsabs.harvard.edu/abs/2023arXiv230503070G}
}

@article{2szy-wypg,
  title = {Two-field formalism for a neural network-enhanced non-Gaussianity search with halos},
  author = {Kvasiuk, Yurii and M\"unchmeyer, Moritz and Smith, Kendrick},
  journal = {Phys. Rev. D},
  volume = {112},
  issue = {2},
  pages = {023540},
  numpages = {22},
  year = {2025},
  month = {Jul},
  publisher = {American Physical Society},
  doi = {10.1103/2szy-wypg},
  url = {https://link.aps.org/doi/10.1103/2szy-wypg}
}

@ARTICLE{Green:2023uyz,
       author = {{Green}, Daniel and {Guo}, Yi and {Han}, Jiashu and {Wallisch}, Benjamin},
        title = "{Light fields during inflation from BOSS and future galaxy surveys}",
      journal = {J. Cosmol. Astropart. Phys.},
         year = 2024,
        month = may,
       volume = {2024},
       number = {5},
          eid = {090},
        pages = {090},
          doi = {10.1088/1475-7516/2024/05/090},
archivePrefix = {arXiv},
       eprint = {2311.04882},
 primaryClass = {astro-ph.CO},
       adsurl = {https://ui.adsabs.harvard.edu/abs/2024JCAP...05..090G}
}

@article{Munchmeyer:2018eey,
    author = {M{\"u}nchmeyer, Moritz and Madhavacheril, Mathew S. and Ferraro, Simone and Johnson, Matthew C. and Smith, Kendrick M.},
    title = "{Constraining local non-Gaussianities with kinetic Sunyaev-Zel{\textquoteright}dovich tomography}",
    eprint = "1810.13424",
    archivePrefix = "arXiv",
    primaryClass = "astro-ph.CO",
    doi = "10.1103/PhysRevD.100.083508",
    journal = "Phys. Rev. D",
    volume = "100",
    number = "8",
    pages = "083508",
    year = "2019"
}

@ARTICLE{2024PhRvD.109d3515G,
       author = {{Goldstein}, Samuel and {Philcox}, Oliver H.~E. and {Hill}, J. Colin and {Esposito}, Angelo and {Hui}, Lam},
        title = "{Consistently constraining f$_{NL}$ with the squeezed lensing bispectrum using consistency relations}",
      journal = {\prd},
         year = 2024,
        month = feb,
       volume = {109},
       number = {4},
          eid = {043515},
        pages = {043515},
          doi = {10.1103/PhysRevD.109.043515},
archivePrefix = {arXiv},
       eprint = {2310.12959},
 primaryClass = {astro-ph.CO},
       adsurl = {https://ui.adsabs.harvard.edu/abs/2024PhRvD.109d3515G}
}

@ARTICLE{2022PhRvD.106l3525G,
       author = {{Goldstein}, Samuel and {Esposito}, Angelo and {Philcox}, Oliver H.~E. and {Hui}, Lam and {Hill}, J. Colin and {Scoccimarro}, Roman and {Abitbol}, Maximilian H.},
        title = "{Squeezing f$_{NL}$ out of the matter bispectrum with consistency relations}",
      journal = {\prd},
         year = 2022,
        month = dec,
       volume = {106},
       number = {12},
          eid = {123525},
        pages = {123525},
          doi = {10.1103/PhysRevD.106.123525},
archivePrefix = {arXiv},
       eprint = {2209.06228},
 primaryClass = {astro-ph.CO},
       adsurl = {https://ui.adsabs.harvard.edu/abs/2022PhRvD.106l3525G}
}

@ARTICLE{2020ApJS..250....2V,
       author = {{Villaescusa-Navarro}, Francisco and {Hahn}, ChangHoon and {Massara}, Elena and {Banerjee}, Arka and {Delgado}, Ana Maria and {Ramanah}, Doogesh Kodi and {Charnock}, Tom and {Giusarma}, Elena and {Li}, Yin and {Allys}, Erwan and {Brochard}, Antoine and {Uhlemann}, Cora and {Chiang}, Chi-Ting and {He}, Siyu and {Pisani}, Alice and {Obuljen}, Andrej and {Feng}, Yu and {Castorina}, Emanuele and {Contardo}, Gabriella and {Kreisch}, Christina D. and {Nicola}, Andrina and {Alsing}, Justin and {Scoccimarro}, Roman and {Verde}, Licia and {Viel}, Matteo and {Ho}, Shirley and {Mallat}, Stephane and {Wandelt}, Benjamin and {Spergel}, David N.},
        title = "{The Quijote Simulations}",
      journal = {Astrophys. J. Suppl. Ser.},
         year = 2020,
        month = sep,
       volume = {250},
       number = {1},
          eid = {2},
        pages = {2},
          doi = {10.3847/1538-4365/ab9d82},
archivePrefix = {arXiv},
       eprint = {1909.05273},
 primaryClass = {astro-ph.CO},
       adsurl = {https://ui.adsabs.harvard.edu/abs/2020ApJS..250....2V}
}

@ARTICLE{2017ComAC...4....2P,
       author = {{Potter}, Douglas and {Stadel}, Joachim and {Teyssier}, Romain},
        title = "{PKDGRAV3: beyond trillion particle cosmological simulations for the next era of galaxy surveys}",
      journal = {Computational Astrophysics and Cosmology},
         year = 2017,
        month = may,
       volume = {4},
       number = {1},
          eid = {2},
        pages = {2},
          doi = {10.1186/s40668-017-0021-1},
archivePrefix = {arXiv},
       eprint = {1609.08621},
 primaryClass = {astro-ph.IM},
       adsurl = {https://ui.adsabs.harvard.edu/abs/2017ComAC...4....2P}
}

@ARTICLE{2021MNRAS.508.4017M,
       author = {{Maksimova}, Nina A. and {Garrison}, Lehman H. and {Eisenstein}, Daniel J. and {Hadzhiyska}, Boryana and {Bose}, Sownak and {Satterthwaite}, Thomas P.},
        title = "{ABACUSSUMMIT: a massive set of high-accuracy, high-resolution N-body simulations}",
      journal = {Mon. Not. R. Astron. Soc.},
         year = 2021,
        month = dec,
       volume = {508},
       number = {3},
        pages = {4017-4037},
          doi = {10.1093/mnras/stab2484},
archivePrefix = {arXiv},
       eprint = {2110.11398},
 primaryClass = {astro-ph.CO},
       adsurl = {https://ui.adsabs.harvard.edu/abs/2021MNRAS.508.4017M}
}

@ARTICLE{2013PASP..125..306F,
       author = {{Foreman-Mackey}, Daniel and {Hogg}, David W. and {Lang}, Dustin and {Goodman}, Jonathan},
        title = "{emcee: The MCMC Hammer}",
      journal = {Publ. Astron. Soc. Pac.},
         year = 2013,
        month = mar,
       volume = {125},
       number = {925},
        pages = {306},
          doi = {10.1086/670067},
archivePrefix = {arXiv},
       eprint = {1202.3665},
 primaryClass = {astro-ph.IM},
       adsurl = {https://ui.adsabs.harvard.edu/abs/2013PASP..125..306F}
}

@misc{DESI:2016fyo,
    author = "Aghamousa, Amir and others",
    collaboration = "DESI",
    title = "{The DESI Experiment Part I: Science,Targeting, and Survey Design}",
    eprint = "1611.00036",
    archivePrefix = "arXiv",
    primaryClass = "astro-ph.IM",
    reportNumber = "FERMILAB-PUB-16-517-AE",
    month = "10",
    year = "2016"
}

@article{EuclidTheoryWorkingGroup:2012gxx,
    author = "Amendola, Luca and others",
    collaboration = "Euclid Theory Working Group",
    title = "{Cosmology and fundamental physics with the Euclid satellite}",
    eprint = "1206.1225",
    archivePrefix = "arXiv",
    primaryClass = "astro-ph.CO",
    doi = "10.12942/lrr-2013-6",
    journal = "Living Rev. Rel.",
    volume = "16",
    pages = "6",
    year = "2013"
}

@misc{SPHEREx:2014bgr,
    author = "Dor{\'e}, Olivier and others",
    collaboration = "SPHEREx",
    title = "{Cosmology with the SPHEREX All-Sky Spectral Survey}",
    eprint = "1412.4872",
    archivePrefix = "arXiv",
    primaryClass = "astro-ph.CO",
    month = "12",
    year = "2014"
}

@misc{LSSTScience:2009jmu,
    author = "Abell, Paul A. and others",
    collaboration = "LSST Science, LSST Project",
    title = "{LSST Science Book, Version 2.0}",
    eprint = "0912.0201",
    archivePrefix = "arXiv",
    primaryClass = "astro-ph.IM",
    reportNumber = "FERMILAB-TM-2495-A, SLAC-R-1031",
    doi = "10.2172/1156415",
    month = "12",
    year = "2009"
}

@ARTICLE{2025OJAp....8E...6H,
       author = {{Harscouet}, Lea and {Cowell}, Jessica A. and {Ereza}, Julia and {Alonso}, David and {Camacho}, Hugo and {Nicola}, Andrina and {Slosar}, An{\v{z}}e},
        title = "{Fast Projected Bispectra: the filter-square approach}",
      journal = {The Open Journal of Astrophysics},
         year = 2025,
        month = jan,
       volume = {8},
          eid = {6},
        pages = {6},
          doi = {10.33232/001c.128309},
archivePrefix = {arXiv},
       eprint = {2409.07980},
 primaryClass = {astro-ph.CO},
       adsurl = {https://ui.adsabs.harvard.edu/abs/2025OJAp....8E...6H}
}

@ARTICLE{Chen:2024exy,
       author = {{Chen}, Xinyi and {Padmanabhan}, Nikhil and {Eisenstein}, Daniel J.},
        title = "{Probing primordial non-Gaussianity by reconstructing the initial conditions}",
      journal = {J. Cosmol. Astropart. Phys.},
         year = 2025,
        month = aug,
       volume = {2025},
       number = {8},
          eid = {055},
        pages = {055},
          doi = {10.1088/1475-7516/2025/08/055},
archivePrefix = {arXiv},
       eprint = {2412.00968},
 primaryClass = {astro-ph.CO},
       adsurl = {https://ui.adsabs.harvard.edu/abs/2025JCAP...08..055C}
}

@article{Seljak:2008xr,
    author = "Seljak, Uros",
    title = "{Extracting primordial non-gaussianity without cosmic variance}",
    eprint = "0807.1770",
    archivePrefix = "arXiv",
    primaryClass = "astro-ph",
    doi = "10.1103/PhysRevLett.102.021302",
    journal = "Phys. Rev. Lett.",
    volume = "102",
    pages = "021302",
    year = "2009"
}

@article{Schmittfull:2017ffw,
    author = "Schmittfull, Marcel and Seljak, Uros",
    title = "{Parameter constraints from cross-correlation of CMB lensing with galaxy clustering}",
    eprint = "1710.09465",
    archivePrefix = "arXiv",
    primaryClass = "astro-ph.CO",
    doi = "10.1103/PhysRevD.97.123540",
    journal = "Phys. Rev. D",
    volume = "97",
    number = "12",
    pages = "123540",
    year = "2018"
}

@ARTICLE{2003JHEP...05..013M,
       author = {{Maldacena}, Juan},
        title = "{Non-gaussian features of primordial fluctuations in single field inflationary models}",
      journal = {Journal of High Energy Physics},
         year = 2003,
        month = may,
       volume = {2003},
       number = {5},
          eid = {013},
        pages = {013},
          doi = {10.1088/1126-6708/2003/05/013},
archivePrefix = {arXiv},
       eprint = {astro-ph/0210603},
 primaryClass = {astro-ph},
       adsurl = {https://ui.adsabs.harvard.edu/abs/2003JHEP...05..013M}
}

@ARTICLE{2004JCAP...10..006C,
       author = {{Creminelli}, Paolo and {Zaldarriaga}, Matias},
        title = "{A single-field consistency relation for the three-point function}",
      journal = {J. Cosmol. Astropart. Phys.},
         year = 2004,
        month = oct,
       volume = {2004},
       number = {10},
          eid = {006},
        pages = {006},
          doi = {10.1088/1475-7516/2004/10/006},
archivePrefix = {arXiv},
       eprint = {astro-ph/0407059},
 primaryClass = {astro-ph},
       adsurl = {https://ui.adsabs.harvard.edu/abs/2004JCAP...10..006C}
}

@ARTICLE{2011JCAP...11..038C,
       author = {{Creminelli}, Paolo and {D'Amico}, Guido and {Musso}, Marcello and {Nore{\~n}a}, Jorge},
        title = "{The (not so) squeezed limit of the primordial 3-point function}",
      journal = {J. Cosmol. Astropart. Phys.},
         year = 2011,
        month = nov,
       volume = {2011},
       number = {11},
          eid = {038},
        pages = {038},
          doi = {10.1088/1475-7516/2011/11/038},
archivePrefix = {arXiv},
       eprint = {1106.1462},
 primaryClass = {astro-ph.CO},
       adsurl = {https://ui.adsabs.harvard.edu/abs/2011JCAP...11..038C}
}

@ARTICLE{2013PhRvD..88h3502P,
       author = {{Pajer}, Enrico and {Schmidt}, Fabian and {Zaldarriaga}, Matias},
        title = "{The Observed squeezed limit of cosmological three-point functions}",
      journal = {\prd},
         year = 2013,
        month = oct,
       volume = {88},
       number = {8},
          eid = {083502},
        pages = {083502},
          doi = {10.1103/PhysRevD.88.083502},
archivePrefix = {arXiv},
       eprint = {1305.0824},
 primaryClass = {astro-ph.CO},
       adsurl = {https://ui.adsabs.harvard.edu/abs/2013PhRvD..88h3502P}
}

@ARTICLE{2019BAAS...51c.107M,
       author = {{Meerburg}, Pieter Daniel and {Green}, Daniel and {Flauger}, Raphael and {Wallisch}, Benjamin and {Marsh}, M.~C. David and {Pajer}, Enrico and {Goon}, Garret and {Dvorkin}, Cora and {Dizgah}, Azadeh Moradinezhad and {Baumann}, Daniel and {Pimentel}, Guilherme L. and {Foreman}, Simon and {Silverstein}, Eva and {Chisari}, Elisa and {Wandelt}, Benjamin and {Loverde}, Marilena and {Slosar}, Anze},
        title = "{Primordial Non-Gaussianity}",
      journal = {Bull. Am. Astron. Soc.},
         year = 2019,
        month = may,
       volume = {51},
       number = {3},
          eid = {107},
        pages = {107},
          doi = {10.48550/arXiv.1903.04409},
archivePrefix = {arXiv},
       eprint = {1903.04409},
 primaryClass = {astro-ph.CO},
       adsurl = {https://ui.adsabs.harvard.edu/abs/2019BAAS...51c.107M}
}

@ARTICLE{2022arXiv220308128A,
       author = {{Ach{\'u}carro}, Ana and {Biagetti}, Matteo and {Braglia}, Matteo and {Cabass}, Giovanni and {Caldwell}, Robert and {Castorina}, Emanuele and {Chen}, Xingang and {Coulton}, William and {Flauger}, Raphael and {Fumagalli}, Jacopo and {Ivanov}, Mikhail M. and {Lee}, Hayden and {Maleknejad}, Azadeh and {Meerburg}, P. Daniel and {Moradinezhad Dizgah}, Azadeh and {Palma}, Gonzalo A. and {Pimentel}, Guilherme L. and {Renaux-Petel}, S{\'e}bastien and {Wallisch}, Benjamin and {Wandelt}, Benjamin D. and {Witkowski}, Lukas T. and {Kimmy Wu}, W.~L.},
        title = "{Inflation: Theory and Observations}",
      journal = {arXiv e-prints},
         year = 2022,
        month = mar,
          eid = {arXiv:2203.08128},
        pages = {arXiv:2203.08128},
          doi = {10.48550/arXiv.2203.08128},
archivePrefix = {arXiv},
       eprint = {2203.08128},
 primaryClass = {astro-ph.CO},
       adsurl = {https://ui.adsabs.harvard.edu/abs/2022arXiv220308128A}
}

@article{Planck:2019kim,
    author = "Akrami, Y. and others",
    collaboration = "Planck",
    title = "{Planck 2018 results. IX. Constraints on primordial non-Gaussianity}",
    eprint = "1905.05697",
    archivePrefix = "arXiv",
    primaryClass = "astro-ph.CO",
    doi = "10.1051/0004-6361/201935891",
    journal = "Astron. Astrophys.",
    volume = "641",
    pages = "A9",
    year = "2020"
}

@ARTICLE{2008PhRvD..77l3514D,
       author = {{Dalal}, Neal and {Dor{\'e}}, Olivier and {Huterer}, Dragan and {Shirokov}, Alexander},
        title = "{Imprints of primordial non-Gaussianities on large-scale structure: Scale-dependent bias and abundance of virialized objects}",
      journal = {\prd},
         year = 2008,
        month = jun,
       volume = {77},
       number = {12},
          eid = {123514},
        pages = {123514},
          doi = {10.1103/PhysRevD.77.123514},
archivePrefix = {arXiv},
       eprint = {0710.4560},
 primaryClass = {astro-ph},
       adsurl = {https://ui.adsabs.harvard.edu/abs/2008PhRvD..77l3514D}
}

@ARTICLE{2008ApJ...677L..77M,
       author = {{Matarrese}, Sabino and {Verde}, Licia},
        title = "{The Effect of Primordial Non-Gaussianity on Halo Bias}",
      journal = {Astrophys. J. Lett.},
         year = 2008,
        month = apr,
       volume = {677},
       number = {2},
        pages = {L77},
          doi = {10.1086/587840},
archivePrefix = {arXiv},
       eprint = {0801.4826},
 primaryClass = {astro-ph},
       adsurl = {https://ui.adsabs.harvard.edu/abs/2008ApJ...677L..77M}
}

@ARTICLE{2008JCAP...08..031S,
       author = {{Slosar}, An{\v{z}}e and {Hirata}, Christopher and {Seljak}, Uro{\v{s}} and {Ho}, Shirley and {Padmanabhan}, Nikhil},
        title = "{Constraints on local primordial non-Gaussianity from large scale structure}",
      journal = {J. Cosmol. Astropart. Phys.},
         year = 2008,
        month = aug,
       volume = {2008},
       number = {8},
          eid = {031},
        pages = {031},
          doi = {10.1088/1475-7516/2008/08/031},
archivePrefix = {arXiv},
       eprint = {0805.3580},
 primaryClass = {astro-ph},
       adsurl = {https://ui.adsabs.harvard.edu/abs/2008JCAP...08..031S}
}

@ARTICLE{2009MNRAS.396...85D,
       author = {{Desjacques}, Vincent and {Seljak}, Uro{\v{s}} and {Iliev}, Ilian T.},
        title = "{Scale-dependent bias induced by local non-Gaussianity: a comparison to N-body simulations}",
      journal = {Mon. Not. R. Astron. Soc.},
         year = 2009,
        month = jun,
       volume = {396},
       number = {1},
        pages = {85-96},
          doi = {10.1111/j.1365-2966.2009.14721.x},
archivePrefix = {arXiv},
       eprint = {0811.2748},
 primaryClass = {astro-ph},
       adsurl = {https://ui.adsabs.harvard.edu/abs/2009MNRAS.396...85D}
}

@ARTICLE{2021JCAP...12..049S,
       author = {{Sailer}, Noah and {Castorina}, Emanuele and {Ferraro}, Simone and {White}, Martin},
        title = "{Cosmology at high redshift - a probe of fundamental physics}",
      journal = {J. Cosmol. Astropart. Phys.},
         year = 2021,
        month = dec,
       volume = {2021},
       number = {12},
          eid = {049},
        pages = {049},
          doi = {10.1088/1475-7516/2021/12/049},
archivePrefix = {arXiv},
       eprint = {2106.09713},
 primaryClass = {astro-ph.CO},
       adsurl = {https://ui.adsabs.harvard.edu/abs/2021JCAP...12..049S}
}

@ARTICLE{2023MNRAS.520.5746A,
       author = {{Andrews}, Adam and {Jasche}, Jens and {Lavaux}, Guilhem and {Schmidt}, Fabian},
        title = "{Bayesian field-level inference of primordial non-Gaussianity using next-generation galaxy surveys}",
      journal = {Mon. Not. R. Astron. Soc.},
         year = 2023,
        month = apr,
       volume = {520},
       number = {4},
        pages = {5746-5763},
          doi = {10.1093/mnras/stad432},
archivePrefix = {arXiv},
       eprint = {2203.08838},
 primaryClass = {astro-ph.CO},
       adsurl = {https://ui.adsabs.harvard.edu/abs/2023MNRAS.520.5746A}
}

@ARTICLE{2020MNRAS.498..464F,
       author = {{Friedrich}, Oliver and {Uhlemann}, Cora and {Villaescusa-Navarro}, Francisco and {Baldauf}, Tobias and {Manera}, Marc and {Nishimichi}, Takahiro},
        title = "{Primordial non-Gaussianity without tails - how to measure f$_{NL}$ with the bulk of the density PDF}",
      journal = {Mon. Not. R. Astron. Soc.},
         year = 2020,
        month = oct,
       volume = {498},
       number = {1},
        pages = {464-483},
          doi = {10.1093/mnras/staa2160},
archivePrefix = {arXiv},
       eprint = {1912.06621},
 primaryClass = {astro-ph.CO},
       adsurl = {https://ui.adsabs.harvard.edu/abs/2020MNRAS.498..464F}
}

@ARTICLE{2021JCAP...04..061B,
       author = {{Biagetti}, Matteo and {Cole}, Alex and {Shiu}, Gary},
        title = "{The Persistence of Large Scale Structures I: Primordial non-Gaussianity}",
      journal = {J. Cosmol. Astropart. Phys.},
         year = 2021,
        month = apr,
       volume = {2021},
       number = {4},
          eid = {061},
        pages = {061},
          doi = {10.1088/1475-7516/2021/04/061},
archivePrefix = {arXiv},
       eprint = {2009.04819},
 primaryClass = {astro-ph.CO},
       adsurl = {https://ui.adsabs.harvard.edu/abs/2021JCAP...04..061B}
}

@article{smith2012halo,
  title={Halo clustering and gNL-type primordial non-Gaussianity},
  author={Smith, Kendrick M and Ferraro, Simone and LoVerde, Marilena},
  journal={Journal of Cosmology and Astroparticle Physics},
  volume={2012},
  number={03},
  pages={032},
  year={2012},
  publisher={IOP Publishing}
}

@ARTICLE{2023PhRvD.107f1301G,
       author = {{Giri}, Utkarsh and {M{\"u}nchmeyer}, Moritz and {Smith}, Kendrick M.},
        title = "{Robust neural network-enhanced estimation of local primordial non-Gaussianity}",
      journal = {\prd},
         year = 2023,
        month = mar,
       volume = {107},
       number = {6},
          eid = {L061301},
        pages = {L061301},
          doi = {10.1103/PhysRevD.107.L061301},
archivePrefix = {arXiv},
       eprint = {2205.12964},
 primaryClass = {astro-ph.CO},
       adsurl = {https://ui.adsabs.harvard.edu/abs/2023PhRvD.107f1301G}
}

@ARTICLE{2001PhRvD..64d3516C,
       author = {{Cooray}, Asantha},
        title = "{Squared temperature-temperature power spectrum as a probe of the CMB bispectrum}",
      journal = {\prd},
         year = 2001,
        month = aug,
       volume = {64},
       number = {4},
          eid = {043516},
        pages = {043516},
          doi = {10.1103/PhysRevD.64.043516},
archivePrefix = {arXiv},
       eprint = {astro-ph/0105415},
 primaryClass = {astro-ph},
       adsurl = {https://ui.adsabs.harvard.edu/abs/2001PhRvD..64d3516C}
}

@ARTICLE{2014PhRvD..89l3522V,
       author = {{Valageas}, Patrick},
        title = "{Angular-averaged consistency relations of large-scale structures}",
      journal = {\prd},
         year = 2014,
        month = jun,
       volume = {89},
       number = {12},
          eid = {123522},
        pages = {123522},
          doi = {10.1103/PhysRevD.89.123522},
archivePrefix = {arXiv},
       eprint = {1311.4286},
 primaryClass = {astro-ph.CO},
       adsurl = {https://ui.adsabs.harvard.edu/abs/2014PhRvD..89l3522V}
}

@ARTICLE{2014PhRvD..90b3546N,
       author = {{Nishimichi}, Takahiro and {Valageas}, Patrick},
        title = "{Testing the equal-time angular-averaged consistency relation of the gravitational dynamics in N-body simulations}",
      journal = {\prd},
         year = 2014,
        month = jul,
       volume = {90},
       number = {2},
          eid = {023546},
        pages = {023546},
          doi = {10.1103/PhysRevD.90.023546},
archivePrefix = {arXiv},
       eprint = {1402.3293},
 primaryClass = {astro-ph.CO},
       adsurl = {https://ui.adsabs.harvard.edu/abs/2014PhRvD..90b3546N}
}

@ARTICLE{2004MNRAS.348..897T,
       author = {{Takada}, Masahiro and {Jain}, Bhuvnesh},
        title = "{Cosmological parameters from lensing power spectrum and bispectrum tomography}",
      journal = {Mon. Not. R. Astron. Soc.},
         year = 2004,
        month = mar,
       volume = {348},
       number = {3},
        pages = {897-915},
          doi = {10.1111/j.1365-2966.2004.07410.x},
archivePrefix = {arXiv},
       eprint = {astro-ph/0310125},
 primaryClass = {astro-ph},
       adsurl = {https://ui.adsabs.harvard.edu/abs/2004MNRAS.348..897T}
}

@misc{LSSTDarkEnergyScience:2018jkl,
    author = "Mandelbaum, Rachel and others",
    collaboration = "LSST Dark Energy Science",
    title = "{The LSST Dark Energy Science Collaboration (DESC) Science Requirements Document}",
    eprint = "1809.01669",
    archivePrefix = "arXiv",
    primaryClass = "astro-ph.CO",
    reportNumber = "FERMILAB-PUB-18-465-A",
    doi = "10.2172/1471560",
    month = "9",
    year = "2018"
}

@BOOK{1988qtam.book.....V,
       author = {{Varshalovich}, D.~A. and {Moskalev}, A.~N. and {Khersonskii}, V.~K.},
        title = "{Quantum Theory of Angular Momentum}",
         year = 1988,
          doi = {10.1142/0270},
       adsurl = {https://ui.adsabs.harvard.edu/abs/1988qtam.book.....V}
}

@ARTICLE{2005ApJ...622..759G,
   author = {{G{\'o}rski}, K.~M. and {Hivon}, E. and {Banday}, A.~J. and 
	{Wandelt}, B.~D. and {Hansen}, F.~K. and {Reinecke}, M. and 
	{Bartelmann}, M.},
    title = "{HEALPix: A Framework for High-Resolution Discretization and Fast Analysis of Data Distributed on the Sphere}",
  journal = {\apj},
   eprint = {arXiv:astro-ph/0409513},
     year = 2005,
    month = apr,
   volume = 622,
    pages = {759-771},
      doi = {10.1086/427976},
   adsurl = {http://adsabs.harvard.edu/abs/2005ApJ...622..759G}
}

@article{Zonca2019,
  doi = {10.21105/joss.01298},
  url = {https://doi.org/10.21105/joss.01298},
  year = {2019},
  month = mar,
  publisher = {The Open Journal},
  volume = {4},
  number = {35},
  pages = {1298},
  author = {Andrea Zonca and Leo Singer and Daniel Lenz and Martin Reinecke and Cyrille Rosset and Eric Hivon and Krzysztof Gorski},
  title = {healpy: equal area pixelization and spherical harmonics transforms for data on the sphere in Python},
  journal = {Journal of Open Source Software}
}

@article{Chang:2013xja,
    author = "Chang, C. and Jarvis, M. and Jain, B. and Kahn, S. M. and Kirkby, D. and Connolly, A. and Krughoff, S. and Peng, E. and Peterson, J. R.",
    title = "{The Effective Number Density of Galaxies for Weak Lensing Measurements in the LSST Project}",
    eprint = "1305.0793",
    archivePrefix = "arXiv",
    primaryClass = "astro-ph.CO",
    reportNumber = "SLAC-PUB-15696",
    doi = "10.1093/mnras/stt1156",
    journal = "Mon. Not. Roy. Astron. Soc.",
    volume = "434",
    pages = "2121",
    year = "2013"
}

@article{Eisenstein:1997ik,
    author = "Eisenstein, Daniel J. and Hu, Wayne",
    title = "{Baryonic features in the matter transfer function}",
    eprint = "astro-ph/9709112",
    archivePrefix = "arXiv",
    reportNumber = "IASSNS-AST-97-51",
    doi = "10.1086/305424",
    journal = "Astrophys. J.",
    volume = "496",
    pages = "605",
    year = "1998"
}

\end{document}